# Reconsidering the design of planar plasmonic lasers: gain, gap layers, and mode competition

Marianne Aellen[1], Aurelio A. Rossinelli[1], Robert C. Keitel[1], Raphael Brechbühler[1], Felipe V. Antolinez[1], Jian Cui[1,2], and David J. Norris[1]

[1]Optical Materials Engineering Lab, Dept. of Mechanical and Process Engineering, ETH Zurich, 8092 Zurich, Switzerland.
[2]Helmholtz Pioneer Campus, Helmholtz Zentrum München, 85764 Neuherberg, Germany.

**Abstract**

Because surface plasmons can be confined below the diffraction limit, metallic lasers that support plasmonic modes can provide miniaturized sources of electromagnetic waves. Such devices often exploit a multilayer design, in which a semiconductor gain layer is placed near a metallic interface with a gap layer in between. However, despite many experimental demonstrations, key considerations for these planar metallic lasers remain understudied, leading to incorrect conclusions about the optimal design. Here, we pursue a detailed experimental and theoretical study of planar metallic lasers to explore the effect of design parameters on the lasing behavior. We print semiconductor nanoplatelets as a gain layer of controllable thickness onto alumina-coated silver films with integrated planar Fabry–Pérot cavities. Lasing behavior is then monitored with spectrally and polarization-resolved far-field imaging. The results are compared with a theoretical waveguide model and a detailed rate-equation model, which consider both plasmonic and photonic modes. We show that the nature of the lasing mode is dictated by the gain-layer thickness. Moreover, by explicitly treating gain in our waveguide model, we find that, contrary to conventional wisdom, a gap layer with high refractive index is advantageous for plasmonic lasing. Additionally, our rate-equation model reveals a regime where plasmonic and photonic modes compete within the same device, raising the possibility of facile, active mode switching. These findings provide guidance for future designs of metallic lasers and could lead to on-chip lasers with controlled photonic and plasmonic output, switchable at high speeds.

**Introduction**

The minimum size of a photonic device is set by the diffraction limit. To develop smaller components, electromagnetic waves known as surface plasmon polaritons (or surface plasmons for short) have been explored[1,2]. Surface plasmons involve photons coupled to electronic oscillations at a metal–dielectric interface. Because they allow tighter mode confinement than photons, plasmonic devices can reduce the size of integrated circuits for electromagnetic waves[3-9]. Indeed, plasmonic modulators have recently been demonstrated with impressive performance in a small footprint[10,11]. However, one disadvantage of plasmonic devices is that they can introduce significant loss[12]. Consequently, structures that amplify surface plasmons through stimulated emission have also been developed[13]. This has led to plasmonic lasers (or "spasers")[14,15], which combine gain with a plasmonic resonator.

The first demonstration of lasing in a surface-plasmon mode utilized a semiconductor nanowire (for gain) that was placed on top of a plasmonic metal with a dielectric gap layer in between[16]. Similar designs, but with different materials, have extended operational wavelengths from the ultraviolet to the near-infrared and have further improved performance, including decreased thresholds and room-temperature operation[17-24]. Because these devices were fabricated through the random placement of semiconductor nanostructures (wires, flakes, *etc.*) on a substrate, which can hinder laser integration, methods have been sought to precisely place colloidal semiconductors[25] and to efficiently extract the lasing output[26,27]. Alternatively, epitaxially grown semiconductor stacks have been employed in electrically-pumped metallic nanolasers[28-30], yielding continuous-wave operation at room temperature[31,32]. However, only one of these electrically injected devices[29] was able to operate in a plasmonic mode[15]. Hence, a thorough understanding of how the lasing mode depends on design parameters of metallic lasers is key for further development of plasmonic nanolasers.

In general, the major challenge in plasmonic lasers is the compensation of large metallic losses with semiconductor gain. Researchers have primarily addressed this issue by reducing losses in the metal through improved fabrication[19,33] and by maximizing gain with optimal semiconductor selection[16,21,24]. However, other design factors have been less studied, leading to potentially incorrect assumptions. For example, it is commonly believed that the electric field should be concentrated in a low-refractive-index dielectric gap layer to reduce metallic losses[16,18,21,23,34]. The effect of this choice on the semiconductor gain has not been considered, possibly resulting in sub-optimal designs. Moreover, the thickness of the semiconductor layer has not been explored in detail. This thickness not only influences the overall gain, but it dictates the confinement and, more importantly, the nature of the lasing mode, namely whether a plasmonic or photonic mode is coherently amplified. Indeed, plasmonic lasers are often distinguished from their photonic counterparts by comparing a plasmonic design with an equivalent structure in which the metal has been replaced by a dielectric material[16,18,35]. However, this approach is insufficient for many plasmonic devices whose dimensions allow them to support both plasmonic and photonic modes. The coexistence and competition of these modes in metallic lasers must be systematically investigated for a full understanding of these devices.

Here, we address this by treating the common semiconductor–insulator–metal laser structure with a gain layer of precisely controlled thickness. We select colloidal semiconductor nanoplatelets (NPLs) and alumina for the gain and gap layers, respectively. Combining spectrally and polarization-resolved far-field measurements with a theoretical waveguide model that explicitly includes losses and gain, we show how the gain-layer thickness largely determines whether a device lases in the plasmonic or photonic mode. Furthermore, we establish design rules that allow for comparison of different device designs to determine whether a high- or low-refractive-index gap layer is advantageous for plasmonic lasing. Our measurements and calculations reveal a previously unstudied regime where plasmonic and photonic modes coexist and compete for gain in an unintuitive manner. This behavior, elucidated by a rate-equation model that explicitly takes into account both types of modes, leads to the possibility of switching between plasmonic and photonic lasing within the same device. More generally, our study clarifies the interplay of loss and gain in multilayer metallic lasers and provides broader insights for more complex designs.

**Results**

**Device fabrication**

Our devices were fabricated through a three-step process with fine control over the thickness and quality of the constituent materials. Our cavities consist of two 500-nm-tall silver (Ag) reflectors on a smooth Ag substrate, forming a stable, 10-µm-long Fabry–Pérot resonator (Fig. 1a). The reflector shape is parabolic with a radius of curvature of 20 µm. The low roughness of the Ag, achieved by template stripping[36,37], minimizes propagation losses while the reflectors serve as efficient mirrors for both plasmonic and photonic modes[38,39]. Using atomic layer deposition, the cavities were then coated with 10 nm of alumina, which both serves as the gap layer and prevents Ag degradation. Finally, a 2-µm-wide stripe of colloidal CdSe/Cd$_x$Zn$_{1-x}$S core/shell NPLs[40] was placed between the reflectors using electrohydrodynamic nanoprinting[25]. By tuning the printing parameters, the thickness of the NPL stripe could be precisely controlled.

**Coexisting waveguide modes**

A critical complication of planar, multilayer devices is that they can support both plasmonic and photonic propagating modes, any of which can lase. Using an theoretical multilayer-waveguide model (see Materials and methods), which consists of a semi-infinite Ag layer, a 10-nm alumina layer, a NPL gain layer of variable thickness $d_{NPL}$, and a semi-infinite air layer, we identify the transverse-magnetic (TM) and transverse-electric (TE) modes at a free-space wavelength $\lambda_0 = 635$ nm (Fig. 1b). Below a gain-layer thickness of 150 nm, only the fundamental TM and TE modes exist. The TM mode is plasmonic, localized at the Ag–dielectric interface, and exists even at vanishingly thin gain layers. The TE mode is photonic and ceases to exist below a gain-layer thickness of ~70 nm (Fig. 1c). This "cutoff" is a key characteristic of the photonic mode.

Spectrally and polarization-resolved far-field measurements reveal the existence of plasmonic and photonic modes in our devices. Within our Fabry–Pérot cavities, propagating modes form longitudinal standing waves whose resonances are revealed when we optically excite the NPL stripe with a light-emitting diode (LED). For example, Fig. 1d plots the leakage spectrum for light scattered at the inner-reflector edge for a device at 4 K. The spontaneous-emission spectrum from the colloidal NPLs is superimposed with peaks representing the cold-cavity modes. Whether these peaks originate from plasmonic or photonic modes can be distinguished using a linear polarizer. Plasmonic (photonic) modes scatter at the reflector into photons with an electric-field component parallel (perpendicular) to the cavity long axis ($z$-direction, see schematic in Fig. 1a). These photons are collected by our microscope objective.

The free spectral range (FSR) further confirms that polarization can be used to distinguish between plasmonic and photonic modes, even within the same device. The measured FSR of TM and TE-polarized cavity modes display two distinct regimes that exhibit a trend closely matching calculated values (see Materials and methods; Fig. 1e). Our calculation slightly overestimates the FSR for the photonic mode, perhaps because it does not capture lateral confinement effects. We hypothesize that close to its cutoff, the photonic mode strongly extends toward the edges of our NPL stripes in the transverse in-plane direction. This would introduce a stronger waveguide dispersion that results in a higher group index and, therefore, a decreased FSR compared to the calculated values. Despite this slight discrepancy, our waveguide model captures all salient features of the plasmonic and photonic modes as resolved by polarization in our far-field measurements.

**Lasing from photonic and plasmonic modes**

To characterize lasing, we optically pumped cavities at 4 K with a defocused pulsed laser beam [Fig. 2a, b; lasing at higher temperatures is discussed in the Supplementary Information (SI) Section S1]. As above, we collect the leakage spectrum of light scattered at the inner-reflector edge of our cavities. At low excitation fluences, broad-band spontaneous emission dominates the cavity spectrum. With increasing pump fluence, distinct lasing peaks appear. As previously shown for the cold-cavity spectra, polarization can be used to resolve which lasing peaks represent photonic and plasmonic modes. In a device with a thin gain layer ($d_{NPL}$ = 56 nm, Fig. 2c), the lasing emission is TM-polarized—indicative of plasmonic lasing. For a thicker gain layer ($d_{NPL}$ = 74 nm, Fig. 2d), photonic (TE-polarized) lasing is observed. Even though the spectra of these comparable devices look similar, the underlying nature of the lasing mode is different and can only be distinguished through polarization-resolved measurements.

The influence of the gain-layer thickness on the lasing mode could be assessed by probing twelve devices with variable NPL-stripe thicknesses. All devices were fabricated on the same substrate to eliminate potential variations in quality or thickness of the Ag and alumina. Their lasing modes were assigned based on their out-scattered polarization. For all devices, the intensity of the lasing peak under one polarization direction was at least one order of magnitude greater than in the orthogonal direction. Hence, each $d_{NPL}$ could be associated with lasing in the plasmonic or the photonic mode, revealing two regimes. For 41 nm ≤ $d_{NPL}$ ≤ 66 nm, plasmonic lasing was observed, while for 74 nm ≤ $d_{NPL}$ ≤ 85 nm, photonic lasing was found (Fig. 2e). For the device with $d_{NPL}$ = 24 nm, lasing could not be achieved, even at the highest pump fluence. While the cold-cavity spectra showed the coexistence of plasmonic and photonic modes for gain layers thicker than the photonic-mode cutoff, only photonic lasing was observed in these devices.

**Modal-gain calculations predict gain-layer requirements**

The dependence on gain-layer thickness can be understood using a modified version of our multilayer-waveguide model. Instead of treating the gain layer as transparent, as is commonly done when optimizing for the modal propagation length[3], we included material gain. The material gain describes the gain per unit length that a plane wave would experience in a uniform infinitely extended medium under a specific set of conditions (excitation density, temperature, *etc.*). To describe the gain experienced by a confined mode in our multilayer-waveguide structure, we then need to determine the modal gain, $G_{mod}$, which is calculated from the imaginary part of the propagation constant, $k_z''$, of the respective mode, through $G_{mod} = -2k_z''$ (ref. 41). The modal gain incorporates both ohmic losses from the Ag and material gain from the gain medium. These are included as positive (loss) or negative (gain) values in the imaginary part of the relative permittivity of the Ag and the gain layers, respectively. The alumina and air are assumed to be lossless with purely real permittivity values.

To achieve lasing, all cavity losses must be compensated by gain. The Ag losses (contained in $G_{mod}$) and reflection losses represent the main loss channels (see SI Section S2), resulting in the following condition for lasing:

$$G_{\text{mod}} \geq -\frac{\ln(R)}{L_{\text{cav}}}, \quad (1)$$

where $R$ is the mirror reflectivity and $L_{cav}$ the cavity length (see SI Section S3 for derivation). The reflectivity of our Ag reflectors is estimated to be ~90% for both plasmonic and photonic modes (see SI Section S4)[38,39]. Hence, lasing in our 10-µm-long cavities can only be obtained for a modal gain equal to or greater than our reflection losses of 105 cm$^{-1}$.

We can calculate the modal gain for a range of material gains, $G_{mat}$, and gain-layer thicknesses, $d_{gain}$, to understand the requirements for lasing in plasmonic and photonic modes. For a transparent gain layer ($G_{mat}$ = 0 cm$^{-1}$, dotted lines in Fig. 3), the modal gain is negative—indicative of dissipative mode propagation due to Ag losses. The plasmonic mode suffers from higher losses than the photonic mode (blue and red dotted lines, respectively, in Fig. 3) for any fixed gain-layer thickness due to the strong mode localization inside the Ag (Fig. 1b). With a material gain of $G_{mat}$ = 1500 cm$^{-1}$ (dashed lines in Fig. 3), the photonic mode compensates propagation and reflection losses when the gain-layer thicknesses is slightly above the photonic-mode cutoff ($d_{gain}$ > 71 nm). This is depicted in Fig. 3 when the red dashed line rises above the horizontal grey line, which represents reflection losses in our cavity. At this thickness, the plasmonic mode is still dominated by losses (blue, dashed line in Fig. 3). This agrees with experiments, where photonic lasing is observed for 74 nm ≤ $d_{NPL}$ ≤ 85 nm. For $G_{mat}$ = 2500 cm$^{-1}$ (solid lines in Fig. 3), the modal gain of the plasmonic mode fulfills the lasing condition for $d_{gain}$ > 41 nm. Our observation of plasmonic lasing for 41 nm ≤ $d_{NPL}$ ≤ 66 nm is completely consistent with these calculations. Furthermore, CdSe NPLs can provide such high gain values[42,43]. Therefore, our multilayer-waveguide model accurately describes the experimental dependence of the lasing mode on the NPL-stripe thickness assuming that a range of material-gain values can be attained upon excitation of the NPL stripe.

Figure 3 also reveals another interesting effect. Even though the photonic mode experiences significantly lower Ag losses, the plasmonic mode displays a larger modal gain than the photonic mode for $G_{mat}$ = 2500 cm$^{-1}$ and 69 nm < $d_{gain}$ < 84 nm (grey shaded area in Fig. 3). We discuss this counter-intuitive result in detail further below. Briefly, the modal gain depends not only on the Ag losses and the material gain but also on the mode confinement within the gain layer. We can express the modal gain in terms of the confinement factor, $\Gamma$, through[41]

$$G_{mod} = \Gamma \cdot G_{mat} - \frac{1}{L_{prop}}, \qquad (2)$$

where $L_{prop}$ is the propagation length of the mode for a transparent gain layer ($G_{mat}$ = 0 cm$^{-1}$). We avoid the complication of finding an analytical expression for the confinement factor[41] by computationally sweeping over the material gain and extracting the confinement factor and propagation length from a linear fit to the modal gain (see SI Section S5). While the plasmonic mode suffers larger losses than the photonic mode (equivalent to having a shorter propagation length, Fig. S4b), its confinement factor is notably larger (Fig. S4c). Therefore, the plasmonic mode achieves a larger modal gain than the photonic mode when the following condition is met:

$$G_{\text{mat}} > \frac{\Delta\alpha_{\text{prop}}}{\Delta\Gamma}, \tag{3}$$

with $\Delta\alpha_{\text{prop}} = (1/L_{\text{prop,TM}} - 1/L_{\text{prop,TE}})$ and $\Delta\Gamma = \Gamma_{\text{TM}} - \Gamma_{\text{TE}}$. According to this mode condition (Eq. 3), the mode with larger $G_{\text{mod}}$ is determined not only by $d_{\text{gain}}$ but also by the exact value of $G_{\text{mat}}$. This effect remains unnoticed if gain is not explicitly included in the mode calculations.

The relevant metric for lasing is the threshold gain—the minimum material gain required to achieve loss compensation. Using the lasing condition (Eq. 1) and the modal-gain expression (Eq. 2), the threshold gain, $G_{\text{th}}$, is defined as:

$$G_{\text{th}} = \frac{1}{\Gamma}\left(\frac{1}{L_{\text{prop}}} - \frac{\ln(R)}{L_{\text{cav}}}\right). \tag{4}$$

The smaller the threshold gain, the lower the required excitation density in the gain medium. This is especially desirable for designs where heating poses an upper limit to the performance (a common complication in nanolasers[31,32]).

The threshold-gain expression (Eq. 4) can explain the lack of lasing in the device with the thinnest NPL stripe. With $d_{\text{gain}} = 24$ nm, we calculate $G_{\text{th}} > 4000$ cm$^{-1}$ (Fig. S4d). Such a high $G_{\text{mat}}$ can only be achieved under extremely high pump fluences[42,43]. Unfortunately, the emission intensity of our NPL stripes decreased irreversibly when the pump fluence was too high, presumably due to detrimental heating effects. Therefore, the threshold gain (and thus lasing) could not be attained. We note that Eq. 4 applies to any mode in any Fabry–Pérot-cavity laser, including more complex two or three-dimensional waveguide systems. This expression not only predicts whether lasing can be achieved in a given mode but allows selection of the appropriate gain material by including material properties that are independent of the design.

**Reconsidering the gap-layer refractive index**

Our multilayer-waveguide model can also be used to analyze the second key component of planar plasmonic lasers—the dielectric gap layer. Because the effect of gain has not been fully treated in the literature, incorrect conclusions about the gap layer can result. In particular, early plasmonic lasers exploited low-index dielectric layers to confine the electric field inside the gap to reduce losses from the metal[16,18,19,21-23,35]. However, experimental results were often compared to calculations of passive (*i.e.* $G_{mat} = 0$ cm$^{-1}$) structures. Thus, the choice of a low-index layer should be revisited to assess whether it is indeed advantageous for plasmonic lasers.

We theoretically compare the plasmonic modes of the same waveguide structure as above, but with the gap layer replaced by either a low-index dielectric, *e.g.* magnesium fluoride ($n_{gap} = 1.42$), or a high-index dielectric, *e.g.* titanium dioxide ($n_{gap} = 2.13$). Assuming a transparent gain layer ($G_{mat} = 0$ cm$^{-1}$), the low-index-gap (LIG) structure achieves larger modal gains than the high-index-gap (HIG) structure for any fixed gain-layer thickness (dashed lines Fig. 4a) due to reduced mode localization inside the Ag layer. This agrees with previous considerations[3]. However, when high material gain is included ($G_{mat} = 2500$ cm$^{-1}$), the opposite can be observed (solid lines Fig. 4a). For thicker gain layers, the HIG structure experiences higher modal gain than the LIG structure and thus can cope with potentially higher reflection losses. This contradicts generally accepted design rules and demands a closer analysis of how the material gain dictates which device configuration experiences a higher modal gain, and ultimately, what material gain is needed to achieve lasing.

We first evaluate the material-gain values that allow a HIG structure to experience a larger modal gain than the LIG structure. Therefore, we use the mode condition in Eq. 3, but instead of comparing the plasmonic and photonic modes of the same structure, we compare the plasmonic modes of the HIG and LIG structures. For low material-gain values ($G_{mat} < 2000$ cm$^{-1}$) or thin gain-layers, the LIG structure experiences a larger modal gain (light-green area in Fig. 4b) than the HIG structure (dark-green area in Fig. 4b). However, for low material-gain values, the relative difference in modal gain only tells us which structure experiences lower ohmic losses, but not whether the mode can lase.

Using the threshold gain (Eq. 4), we can identify the material gain for which loss compensation (and thus lasing) in the plasmonic mode is possible. Assuming a cavity with perfect mirrors (see blue solid and dashed lines labeled $R = 1.0$ in Fig. 4b), the crossing point of the threshold gain for the two structures (HIG and LIG) is at $d_{gain} \approx 43$ nm. For devices with thinner (thicker) gain layers, a high-index (low-index) gap layer is preferred, as the threshold gain is lower for this structure.

Upon introducing reflection losses ($R < 1.0$), the threshold-gain crossing point moves along the mode-condition border (*i.e.* the border between the light- and dark-green areas in Fig. 4b) toward thicker gain layers, extending the range of gain-layer thicknesses for which the HIG structure is preferred. Additionally, the difference in threshold gain between the HIG and LIG structures becomes significantly larger the thinner the gain layer (left of the crossing point). For example, for $R = 0.5$ in a 10-µm long cavity and $d_{gain} = 40$ nm, the LIG-structure threshold gain is 18.4% larger than the HIG-structure threshold gain, an amount that can be decisive in achieving lasing. Therefore, a LIG structure is only advantageous for plasmonic lasers with a high mirror reflectivity and a thick gain layer. However, for thick gain layers, the photonic mode could become the prevailing lasing mode due to lower Ag losses (see cutoff thickness for the photonic mode in Fig. 4b). For plasmonic lasers with gain-layer thicknesses well below the photonic-mode cutoff thickness, the HIG structure exhibits a lower threshold gain and thus requires lower excitation densities to achieve lasing.

Physically, this can be understood in terms of mode localization in the constituent layers. It is indeed the case that Ag losses are reduced for the LIG structure due to localization of the mode inside the gap layer. However, this also diminishes the mode interaction with the gain layer, and ultimately lowers the modal gain. This effect has been overlooked when designing plasmonic lasers. In cavities that employ a semiconductor gain layer with a higher refractive index (*e.g.* $n_{gain} > 2$ instead of NPL films with $n_{NPL} = 1.89$, see SI Section S6), the difference in threshold gain between the HIG and LIG structures is even greater (Fig. S5a, b), showcasing the advantage of high-index gap layers. Similar conclusions were found in two-dimensional waveguide calculations[44]. In principle, removing the gap layer altogether would further reduce the threshold gain (Fig. S5c). However, this renders the Ag prone to degradation. Also, if colloidal nanocrystals are used as the gain medium, direct contact with Ag can detrimentally affect their emission properties[45]. More generally, the role of the gap layer is only adequately judged if gain is explicitly treated in the mode calculations, an insight that also applies to other plasmonic-laser designs.

**Pump-power dependence on lasing dynamics**

So far, we have evaluated design criteria based on a static material gain. However, experimentally, the material gain will be a function of the pump power and thus will be strongly time dependent under pulsed excitation. We first investigate the output intensity as a function of pump fluence (light–light curve) and then model the findings with rate equations that disclose mode-dependent lasing dynamics beyond findings from a static material gain.

We experimentally measured the light–light curves for two devices (one displaying plasmonic and the other photonic lasing) by extracting the integrated intensity of lasing spectra at various pump fluences (filled circles in Fig. 5a, b). The light–light curves look quite different. While the output of the device with $d_{NPL}$ = 56 nm remains nearly linear with input even when lasing begins [recognizable by the decrease in the spectral linewidth (empty circles in Fig. 5a)], the device with $d_{NPL}$ = 74 nm shows a marked inflection at the corresponding lasing threshold. To elucidate this behavior, we set up a rate-equation model (see Materials and methods) with which we calculated the light–light curves (dotted lines in Fig. 5a, b). Without any fitting (apart from a normalization to capture the unknown collection efficiency), our model can reproduce the light–light curves for both experimentally probed devices.

Using our rate-equation model, we can then study the difference in the light–light curves by looking at the excited-carrier population that feeds into the lasing mode above and below threshold. Above threshold, nearly all carriers decay into the lasing mode, while below threshold, only a fraction does (see SI Section S7). This fraction is $\beta_i \Phi$, where $\beta_i$ is the spontaneous-emission factor of the mode $i$ ($i$ = TM or TE) and $\Phi$ is the quantum yield of the gain medium. $\beta_i$ can be calculated from the power dissipated by a position- and orientation-averaged electric point dipole into the respective mode (see SI Section S8). We assume $\Phi$ is 88%, as measured for our NPLs in liquid dispersion[40]. In a waveguide structure with a 10-nm alumina gap layer, $\beta_{TE}$ < 10% for $d_{gain}$ = 74 nm, while $\beta_{TM}$ > 70% for $d_{gain}$ = 56 nm. The large $\beta_{TM}$ is indicative of a highly confined, sub-diffraction mode, as observed in numerous nanolasers[16,46]. These values explain why the device with $d_{NPL}$ = 74 nm exhibits a superlinear increase in the output intensity at the threshold. Only ~8% of the carriers feed into the photonic mode below threshold. In contrast, for the device with $d_{NPL}$ = 56 nm, > 60% of the carriers already decay into the plasmonic mode below threshold. Consequently, the output intensity insignificantly rises upon reaching threshold and the light–light curve remains linear. Therefore, the metallic-cavity lasers studied here, despite only differing in gain-layer thickness, exhibit strikingly different light–light curves due to the different nature of their lasing modes.

Our rate-equation model not only describes these light–light curves, but also the time-dependent lasing dynamics, allowing the mode competition between coexisting waveguide modes to be understood. Returning to the light–light curve for $d_{\text{gain}} = 74$ nm, we can appreciate why photonic lasing is observed in the device even though the modal gain of the plasmonic mode is expected to be larger at $G_{\text{mat}} = 2500$ cm$^{-1}$ (see grey area in Fig. 3). Using the model, we can determine the output-photon number of each mode separately. (For simplicity, we use the term "output photon" for both surface plasmons and photons that are lost through imperfect reflection at the cavity mirrors.) We find that the photonic mode exceeds its threshold at a pump fluence of 15 μJ/cm$^2$ (diamond 1 in Fig. 5c). This is consistent with the photonic laser pulse in the time evolution of the output-photon rate (panel 1 in Fig. 5d). Concurrently, the plasmonic output-photon number decreases because the photonic mode quickly depletes the carrier population through stimulated emission. The threshold gain of the photonic mode (Eq. 4) is reached before the mode condition (Eq. 3) can be fulfilled for the plasmonic mode. Even though the confinement factor is larger for the plasmonic mode, the decreased losses of the photonic mode allow photonic lasing at lower pump fluences. Hence, in agreement with experiments on devices with 74 nm $\leq d_{\text{NPL}} \leq$ 85 nm, the photonic mode always reaches its lasing threshold before potential plasmonic lasing can be observed. This assumes low reflection losses for both modes, which is typical for metallic-cavity lasers (see SI Section S9 for the required difference in reflectivity needed to reach the plasmonic threshold gain first).

Even though the photonic mode is the prevailing lasing mode for a metallic-cavity laser with $d_{\text{NPL}} = 74$ nm, our rate-equation model predicts that plasmonic lasing can be achieved if the material gain is sufficiently high. At a pump fluence of 23 μJ/cm$^2$, the plasmonic mode also exceeds its lasing threshold, as evident from the appearance of a plasmonic laser pulse in the time evolution of the output-photon rate (panel 2 in Fig. 5d). At higher pump fluences, the plasmonic mode begins to outcompete the photonic mode, leading to an inversion of their output-photon numbers. For example, plasmonic lasing dominates at 32 μJ/cm$^2$ (panel 3 in Fig. 5d), effectively giving rise to a switching of the primary lasing mode from photonic to plasmonic within the same device. This behavior originates from the difference in confinement factor (as discussed above). At large pump fluences, the material gain is high, and therefore, the mode condition (Eq. 3) is met, resulting in strong carrier-population depletion by the plasmonic mode, and thus, a reduced photon population. Although our experimentally probed devices displayed signs of degradation when the pump fluence was increased far beyond the photonic lasing threshold, the predicted switching mechanism would in principle result in a new type of laser device that could be used as a single controllable source of coherent photons or coherent surface plasmons.

**Discussion**

Our fabrication approach permits a systematic study of the plasmonic and photonic lasing modes in planar multilayer metallic lasers. The solution processability of colloidal NPLs enables the deposition of a gain layer with a precisely controlled thickness. Furthermore, the open Fabry–Pérot cavity allows us to probe the cavity modes through polarization-resolved spectral imaging in the far-field—an unambiguous method for discriminating plasmonic from photonic modes. The large material gain of NPLs enables plasmonic lasing in a cavity that is sub-diffraction in one dimension. In contrast to previous reports that compared plasmonic lasing on metal substrates to photonic lasing on dielectric substrates, we can investigate both plasmonic and photonic lasing in the same system.

Our multilayer-waveguide model demonstrates the existence of plasmonic and photonic modes and predicts the gain requirements for lasing as a function of the gain-layer thickness. To find suitable designs, we not only examine the metallic losses but also the role of gain. This leads to two simple expressions: the threshold gain (defining the minimal gain required to reach lasing) and the mode condition (revealing the gain at which two modes experience the same modal gain). These expressions are universally applicable to Fabry–Pérot-type cavity lasers and allow a comparison of different modes for various material combinations. Using these expressions, we investigate semiconductor–insulator–metal structures with a high- and low-index gap layer. In contrast to the common assumption that a low-index-dielectric gap facilitates plasmonic lasing, we show that for plasmonic lasers with a gain-layer thickness well below the photonic-mode cutoff thickness, a high-index dielectric should be employed. The same conclusion can apply to more complex structures, such as those employed in electrically injected nanolasers[29].

Finally, we use a time-resolved rate-equation model to calculate the pump-power-dependent output characteristics of the plasmonic and photonic lasing modes. Without applying any fitting, our model describes the experimental light–light curves. From this model, we further predict a laser device that switches its dominant lasing mode from photonic to plasmonic upon increasing the pump fluence. Such a device is an interesting candidate for a single laser with a switchable output of plasmonic and photonic modes. Since both modes evolve on picosecond timescales, fast switching speeds would be possible. We note that pump-fluence-independent switching schemes may also be possible, for example, by modulating the mirror reflectivity.

Although numerous plasmonic lasers have been demonstrated, our work illustrates that a fundamental understanding of the lasing modes is crucial for further improvements. In particular, current plasmonic lasers often operate in size regimes where photonic modes can coexist. To confine modes below the diffraction limit, lasing in plasmonic modes is essential. Furthermore, by including gain in mode calculations and by understanding the time dynamics we can better select design parameters to achieve lasing in progressively smaller devices. This can also potentially lead to devices with new functionalities, such as control over the type of output mode.

**Materials and methods**

**Fabrication of laser cavities**

Ag cavities were fabricated by template stripping to achieve a smooth Ag surface[37]. The templates were prepared from 1-mm-thick, 2-inch-diameter, <100> silicon wafers. An electron-beam-lithography step (Vistec Lithography, EBPG 5200+) and a subsequent hydrogen-bromide-based inductively-coupled-plasma reactive-ion etch (Oxford Instruments, Plasmalab System 100) resulted in ~500-nm-deep patterns (for the reflectors) in the template (see SI Section S10 for a thorough description of the cavity geometry).

A film of Ag (Kurt J. Lesker, 99.99%) ~700 nm thick was then deposited on the template through thermal evaporation (Kurt J. Lesker, Nano36) at a base pressure of $< 9 \times 10^{-8}$ mbar and at a deposition rate of 25 Å/s while the template was rotated at 60 rpm. A microscope slide was bonded to the Ag film using epoxy (Epoxy Technology, EPO-TEK OG116-31) that was cured under ultraviolet light for ~2 h. The Ag cavities were stripped manually from the template shortly before alumina was deposited at 50 °C *via* 100 atomic-layer-deposition cycles (Picosun, Sunale R-150). The templates were reused multiple times.

The gain layer involved CdSe/Cd$_x$Zn$_{1-x}$S core/shell NPLs with 4-monolayer-thick CdSe cores and 2-nm-thick shells[40]. The ink for the electrohydrodynamic nanoprinting was prepared by transferring the NPLs from hexane to tetradecane through selective evaporation while adjusting the concentration to an optical density of 5.0 (measured at the lowest-energy exciton peak using a quartz cuvette with a 10-mm path length). A description of the nanoprinting setup can be found elsewhere[47]. Printing was performed by applying 250 V direct current between the metal-coated nozzle (+) and the indium-tin-oxide-coated glass sample holder (ground). The 10-µm-long NPL stripes were generated by moving the sample stage in a serpentine-like fashion to print nine parallel lines at a pitch of 250 nm. Different stripe thicknesses were achieved by varying the number of overprints. To determine the required stage velocity and number of overprints to produce a given stripe thickness, a parameter sweep was performed before printing into cavities using the same ink-loaded nozzle. Stripes were printed on a flat Ag–alumina substrate and examined in reflection on a dark-field optical microscope [Nikon Eclipse LV100, 50× objective (Nikon CFI LU Plan Fluor BD) with a numerical aperture of 0.8]. The stripe thickness was estimated by comparing the stripe color to a reference image of stripes of known thicknesses. After all optical measurements were performed, the stripe thickness of each device was measured using atomic force microscopy (Bruker, Dimension FastScan). See SI Section S10 for a more detailed description of the fabrication methods.

**Optical characterization**

All optical measurements were performed in a closed-cycle helium cryostat (Montana Instruments, Cryostation 2 with LWD option) under vacuum and cooled to 4 K. For cold-cavity spectra, the cavities were illuminated by light from a 385-nm LED (Thorlabs, M385LP1). Lasing experiments were performed with 405-nm laser pulses (~340 fs pulse duration, 1 kHz repetition rate) emerging from a collinear optical parametric amplifier (Spectra-Physics, Spirit-OPA) pumped by a 1040-nm laser (Spectra-Physics, Spirit-1040-8). A defocused laser spot of ~30 µm diameter was generated by employing a defocusing lens before the beam was directed through a 60× extra-long-working-distance objective (Nikon, CFI S Plan Fluor ELWD with a numerical aperture of 0.7) to the sample.

The same objective was used to collect emission from the sample. The emission was separated from the excitation through a 405-nm dichroic longpass filter (AHF analysentechnik, F48-403) and further filtered by a 450-nm longpass filter (Thorlabs, FEL0450). Then, the image was relayed into an imaging spectrometer (Andor, Shamrock 303i) with an entrance slit set to 50 μm. The sample was placed such that the inner edge of a reflector for the cavity under examination was imaged on the vertical entrance slit. The spectrometer dispersed the image horizontally using a 300 lines/mm grating (500-nm blaze) and imaged with an air-cooled electron-multiplying charged-coupled-device camera (Andor, iXon 888 Ultra). For polarization-resolved spectra, a linear polarizer (Thorlabs, LPVISB100-MP2) was placed in an image plane after the 450-nm longpass filter. A more detailed description of the optical characterization can be found in SI Section S11.

**Multilayer-waveguide model**

The multilayer-waveguide model was implemented following a theoretical model described in ref. 48 (see SI Section S12 for details). We solved the eigenvalue equation for TM and TE polarized waves by a minimization algorithm. From the resulting propagation constant, $k_z$, we can derive the effective mode index, $n_{\text{eff}}$, the modal gain, $G_{\text{mod}}$, and the electric- and magnetic-field profiles. As input into the model, the free-space wavelength was set to $\lambda_0 = 635$ nm (1.95 eV) for all calculations except for the FSR calculations, where $\lambda_0$ was varied over a range of 500–700 nm (1.48–1.77 eV). The relative permittivity of each constituent material was obtained from ellipsometry (see SI Section S13).

**Free-spectral-range calculation**

The FSR was calculated with[49]

$$E_{\text{FSR}} = \frac{hc}{2 \cdot L_{\text{cav}} \cdot n_{\text{g}}}, \tag{5}$$

where $h$ is Planck's constant, $c$ the speed of light, and $n_g$ the group index. The group index was calculated from the effective mode index and its dispersion (see SI Section S14) using

$$n_{\text{g}} = n_{\text{eff}} - \lambda \frac{\partial n_{\text{eff}}}{\partial \lambda} \tag{6}$$

with $\lambda$ being the wavelength. The mode index was obtained from the multilayer-waveguide model, where the absorption in the NPL layer was taken into account as obtained from ellipsometry (see SI Section S13).

**Modal-gain calculation**

For the modal-gain calculation, the material gain of the gain layer was swept over a range of values. Therefore, the imaginary part of the relative permittivity was set to negative values, while the real part was kept constant. For a given $G_{\text{mat}}$, the corresponding imaginary part of the relative permittivity, $\varepsilon''_{\text{gain}}$, was calculated using[41]

$$\varepsilon''_{\text{gain}} = -\frac{G_{\text{mat}} n_{\text{NPL}}}{k_0} \tag{7}$$

with $n_{\text{NPL}}$ being the real part of the refractive index of the NPL film as measured from ellipsometry (see SI Section S13) and $k_0 = 2\pi/\lambda_0$ being the wavevector of the free-space wavelength $\lambda_0$.

**Laser rate equations**

To model the power-dependent laser characteristics, we used a coupled rate-equation model[49,50] that describes an excited-carrier population, and two modal populations (surface plasmon and photon) or one modal population (surface plasmon) for laser cavities with a gain-layer thickness above or below the photonic-mode cutoff, respectively (see SI Section S15 for details). The carrier population is fed by a time-dependent pump pulse [and reabsorption (corresponding to negative gain)] and decays through spontaneous emission (radiative and non-radiative) and stimulated emission. The surface-plasmon and photon populations grow through spontaneous and stimulated emission and are depleted by Ag and reflection losses (and reabsorption). All model parameters were estimated from experimental data or retrieved from literature reports. The rate equations were solved through numerical integration for a range of pump fluences. The output-photon number was obtained by integrating the mirror loss rate over time.

**Acknowledgements**

We thank J. Faist and A. Cocina for stimulating discussions and S. Meyer, H. Rojo, R. Grundbacher, U. Drechsler, and A. Olziersky for technical assistance. We are grateful to T. Lendenmann, P. Rohner, and D. Poulikakos from the Laboratory of Thermodynamics in Emerging Technologies at ETH Zurich for assistance with and use of their electrohydrodynamic-nanoprinting setup. This work was supported the European Research Council under the European Union's Seventh Framework Program (FP/2007-2013)

ERC Grant Agreement Nr. 339905 (QuaDoPS Advanced Grant) and by the Swiss National Science Foundation under Award No. 200021-165559.

**Author contributions**

M.A., J.C., and D.J.N. conceived the ideas and planned the experiments. M.A. designed, fabricated, and optically characterized the laser devices with assistance from R.C.K. and R.B. A.A.R. synthesized the nanoplatelets. M.A. performed all calculations with inputs from R.B. and F.V.A. The manuscript was written by M.A., J.C., and D.J.N. with inputs from all authors. D.J.N. supervised the project.

**Conflict of interest**

The authors declare no competing financial interests.

**Data availability**

The data that support the findings of this study are available from the corresponding author on reasonable request.

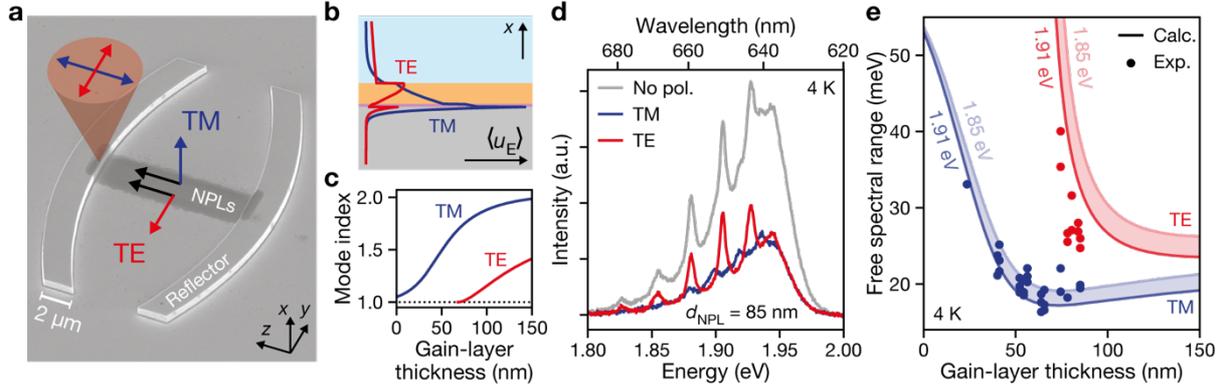

**Fig. 1. Laser device and waveguide model. a** Scanning-electron-microscope image of a laser device (tilted view). The Ag reflectors protrude from a smooth, flat Ag surface. A stripe of NPLs is printed into the center of the cavity. Plasmonic and photonic cavity modes with transverse-magnetic (TM, blue) and transverse-electric (TE, red) field components can be distinguished through the polarization of the leakage radiation that is out-scattered at the inner edge of the reflector, as shown. The black arrows denote the propagation direction. The coordinate system indicates the surface normal of the planar multilayer-waveguide model ($x$) and the two in-plane directions ($y$ and $z$) from which $z$ is chosen as propagation direction. **b** Calculated mode profiles along the surface normal ($x$) of the TM (plasmonic, blue) and TE (photonic, red) modes in a planar multilayer-waveguide structure at a free-space wavelength of $\lambda_0 = 635$ nm. The mode profiles are plotted as the time-averaged electric energy density, $\langle u_E \rangle$. For each mode, $\langle u_E \rangle$ is normalized by the total energy, $U_{EM}$, in the mode. The layers are: Ag (grey), 10-nm-thick alumina (purple), 74-nm-thick gain layer (orange), and air (light blue). **c** Effective mode indices of the TM (plasmonic, blue) and TE (photonic, red) modes as a function of gain-layer thickness calculated for a free-space wavelength $\lambda_0 = 635$ nm. The photonic mode ceases to exist below ~70 nm. **d** Polarization-resolved cold-cavity spectra under illumination from a 385-nm LED at 4 K. The unpolarized emission (grey) is composed of cavity modes originating from plasmonic (TM, blue) and photonic (TE, red) propagating modes that are resolved by analyzing the orthogonal polarizations individually. **e** Comparison of the calculated (shaded areas between lines) and experimentally measured (filled circles) FSRs for 1.85–1.91 eV for the plasmonic (blue) and photonic (red) modes. The experimental data points represent various cavities probed at 4 K under LED illumination.

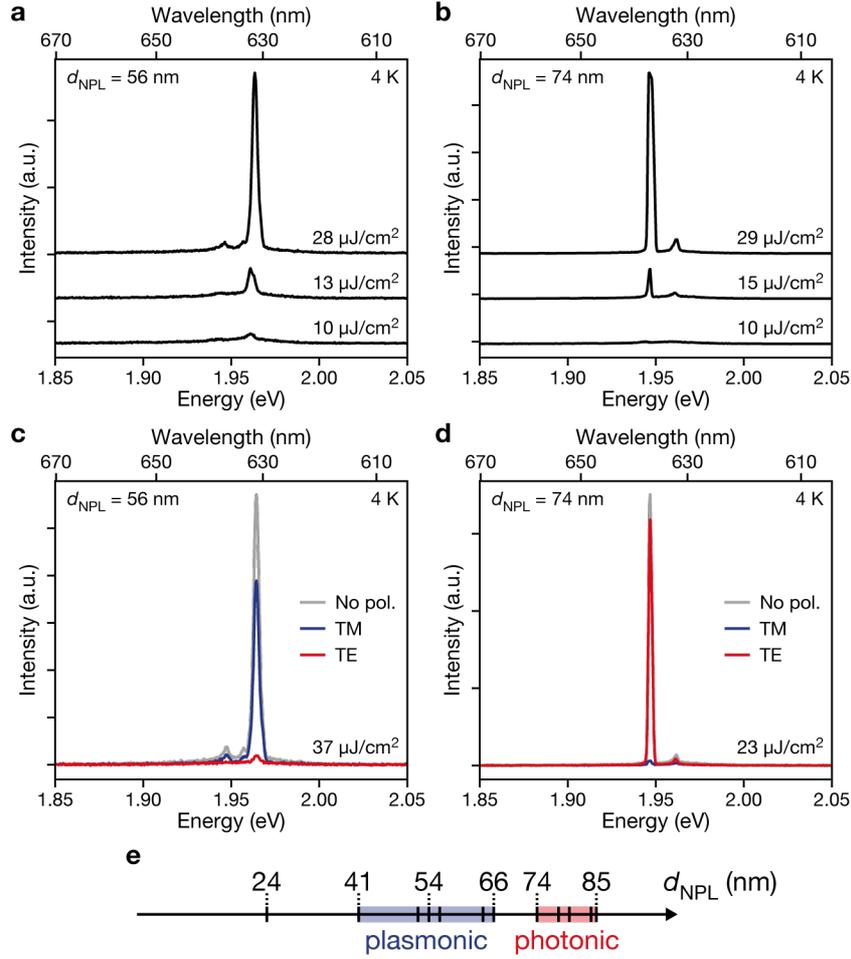

**Fig. 2. Lasing behavior as a function of NPL-stripe thickness. a, b** Laser device emission spectra as a function of pump fluence for two devices with NPL-stripe thicknesses, $d_{NPL}$, of 56 and 74 nm, respectively. The devices were excited at 4 K using a pulsed laser excitation at 405 nm. **c, d** Polarization-resolved lasing spectra reveal the plasmonic (TM, blue) and photonic (TE, red) nature of the lasing emission from **a** and **b**, respectively. **e** Twelve probed devices are marked as tick marks on the axis denoting their NPL-stripe thickness. The colored shading indicates which type of mode, plasmonic (blue) or photonic (red), lased in each device. The tick mark at 24 nm, outside of any colored shading, represents a device for which lasing could not be observed.

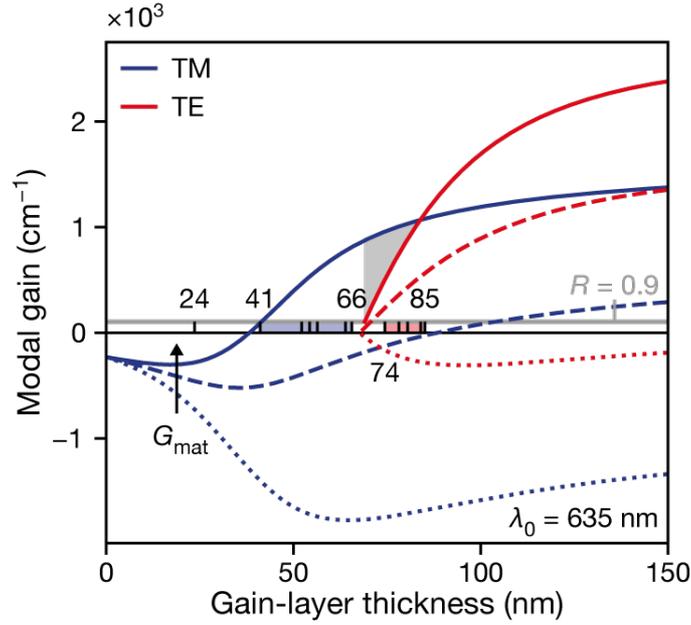

**Fig. 3. Modal gain as a function of gain-layer thickness.** The modal gain for the plasmonic (TM, blue lines) and photonic (TE, red lines) modes at a free-space wavelength $\lambda_0 = 635$ nm is plotted as a function of gain-layer thickness for various material-gain values. The calculations are based on the multilayer-waveguide model. The lines represent material-gain values, $G_{mat}$, of 0 cm$^{-1}$ [corresponding to a transparent gain layer (dotted lines)], 1500 cm$^{-1}$ (dashed lines), and 2500 cm$^{-1}$ (solid lines). The grey horizontal line indicates the reflection losses for a 10-μm-long cavity with mirror reflectivity $R = 0.9$. The grey shaded area marks a regime where the plasmonic modal gain is larger than the photonic modal gain for $G_{mat} = 2500$ cm$^{-1}$. The horizontally distributed tick marks indicate experimentally measured devices (the same as in Fig. 2e) with colored shading marking which type of mode, plasmonic (blue) or photonic (red), lased in each device. For the device at 24 nm, lasing could not be observed.

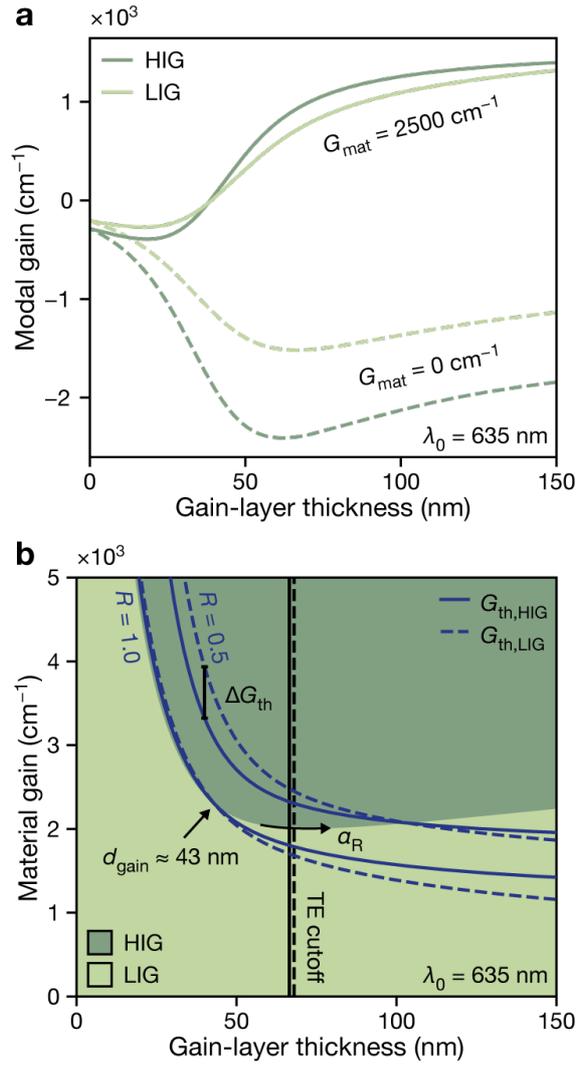

**Fig. 4. Influence of the gap-layer refractive index on the modal gain. a** The modal gain of the plasmonic mode plotted as a function of gain-layer thickness for a device with a NPL gain layer and a 10-nm-thick high-index (HIG; $n$ = 2.13; dark green) or low-index (LIG; $n$ = 1.42; light green) gap layer at a free-space wavelength $\lambda_0$ = 635 nm. The modal gain is plotted for $G_{mat}$ = 0 cm$^{-1}$ (dashed lines) and $G_{mat}$ = 2500 cm$^{-1}$ (solid lines). The structure that experiences a larger modal gain depends on the actual material-gain value. **b** Material gain at which the modal gain is larger for the HIG (dark-green area) or LIG (light-green area) structures for the plasmonic mode. The threshold gains, $G_{th}$, for the plasmonic modes of the HIG and LIG structures are plotted as solid and dashed blue lines, respectively. They are calculated for a 10-μm-long cavity with mirror reflectivities $R$ = 1.0 and 0.5. The arrow at $d_{gain}$ ≈ 43 nm indicates the crossing point of the threshold gains for $R$ = 1.0. For increasing reflection losses, $\alpha_R$, the threshold-gain crossing moves along the mode-condition border in the direction of the arrow. The difference in threshold gain, $\Delta G_{th}$, for $R$ = 0.5 and $d_{gain}$ = 40 nm is 18.4% (vertical bar with caps). The black vertical solid and dashed lines indicate the photonic-mode (TE) cutoff thickness for the HIG and LIG structures, respectively.

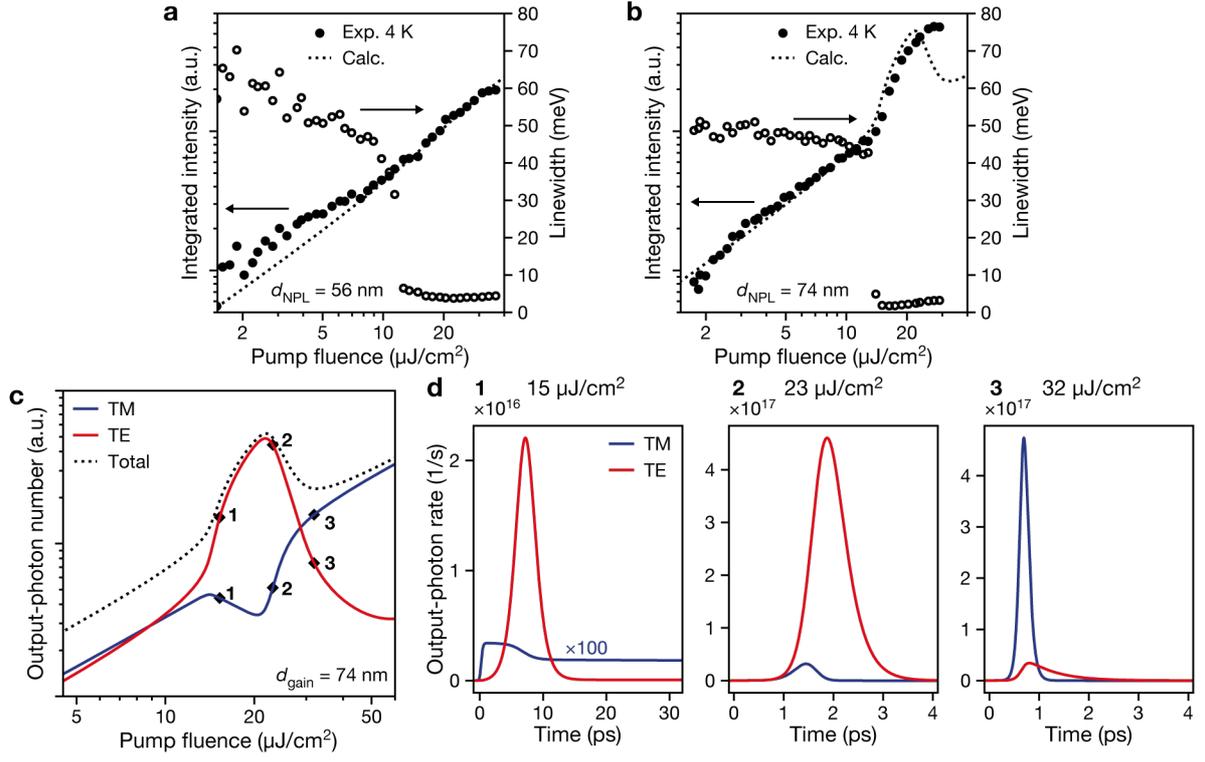

**Fig. 5. Light–light curves and mode switching. a, b** Spectrally integrated output intensity (filled circles) as a function of pump fluence for a device with $d_{NPL}$ = 56 nm (same device as in Fig. 2a, c) and a device with $d_{NPL}$ = 74 nm (same device as in Fig. 2b, d) measured at 4 K under pulsed excitation at 405 nm. Calculated output-photon numbers from the rate-equation model (dotted lines) agree well with the experimentally obtained values. The linewidth of the lasing spectrum (obtained by a fit to a single Gaussian function) indicates the pump-dependent emergence of a narrow lasing peak (empty circles). **c** Calculated output-photon number of the plasmonic (TM, blue) and photonic (TE, red) modes for a cavity with $d_{gain}$ = 74 nm. The dotted line indicates the sum of the two. **d** Time evolution of the output-photon rate for the plasmonic (TM, blue) and photonic (TE, red) modes at three different pump fluences, as indicated by the diamonds in **c**. The primary lasing mode switches from photonic (diamond 1 and 2) to plasmonic (diamond 3) as the pump fluence increases.

Supplementary Information

# Reconsidering the design of planar plasmonic lasers: gain, gap layers, and mode competition

Marianne Aellen[1], Aurelio A. Rossinelli[1], Robert C. Keitel[1], Raphael Brechbühler[1], Felipe V. Antolinez[1], Jian Cui[1,2], and David J. Norris[1]

[1]Optical Materials Engineering Lab, Dept. of Mechanical and Process Engineering, ETH Zurich, 8092 Zurich, Switzerland.
[2]Helmholtz Pioneer Campus, Helmholtz Zentrum München, 85764 Neuherberg, Germany.

## Contents

**Section S1. Temperature dependence**

We have probed our devices at various temperatures ranging from 4 K to room temperature. With increasing temperature, devices with thinner gain layers stopped lasing (Fig. S1). At 100 K, devices thicker than the photonic-mode cutoff thickness were still displaying lasing in the photonic mode, while only the device with the thickest gain layer below the photonic-mode cutoff thickness (device with $d_{NPL}$ = 66 nm) was able to lase in the plasmonic mode (Fig. S1c). At room temperature, only the device with the thickest gain layer ($d_{NPL}$ = 85 nm) displayed photonic lasing (Fig. S1d). It has been reported that the optical-gain threshold for colloidal NPLs is reduced at cryogenic temperatures[S1]. We believe that in our metallic-cavity lasers, we are limited by the amount of power that can be deposited into a device before detrimentally affecting the NPLs possibly due to heating. Hence, we conclude that, at cryogenic temperatures, we are able to achieve larger material gains than at room temperature as we can pump further above the optical-gain threshold of our NPLs before introducing damage. Furthermore, we expect that the losses in the Ag are slightly less at cryogenic temperatures than at room temperature[S2]. The temperature dependence of our metallic-cavity lasers falls in line with the modal gain calculations of our waveguide model.

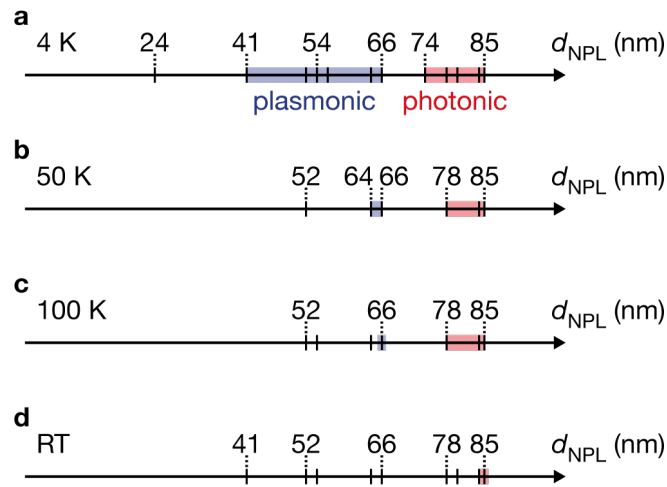

**Fig. S1. Temperature dependence of laser devices.** Multiple NPL laser devices were measured at 4 K (**a**), 50 K (**b**), 100 K (**c**), and room temperature (RT; **d**). Each tick mark on the axis corresponds to a device with given NPL-stripe thickness. The colored shading indicates which type of mode, plasmonic (blue) or photonic (red), lased in each device. Tick marks outside of any colored shading represent devices for which lasing could not be attained.

**Section S2. Major loss channels**

Besides the Ag and reflection losses, scattering could also lead to a loss channel. However, scattering losses are neglected in the lasing condition (Eq. 1 in the main text) because it is expected that the Ag and reflection losses play a more significant role. To verify this assumption, we look at the emission intensity of the NPL stripe and compare it to the intensity arising from the reflector edge when the cavity is pumped above threshold (Fig. S2). The reflector edges light up more brightly than the rest of the stripe. The NPL stripes look very uniform in intensity, suggesting that the NPLs form a smooth film and do not provide scattering centers. Furthermore, from our modal-gain calculations, we find that the Ag losses (corresponding to the negative modal gain for $G_{mat} = 0$ cm$^{-1}$; dotted lines in Fig. 3) are much greater than the reflection losses (grey line in Fig. 3) for most gain-layer thicknesses. Thus, we conclude that scattering losses can be neglected in the lasing condition, whereas Ag and reflection losses should be taken into consideration.

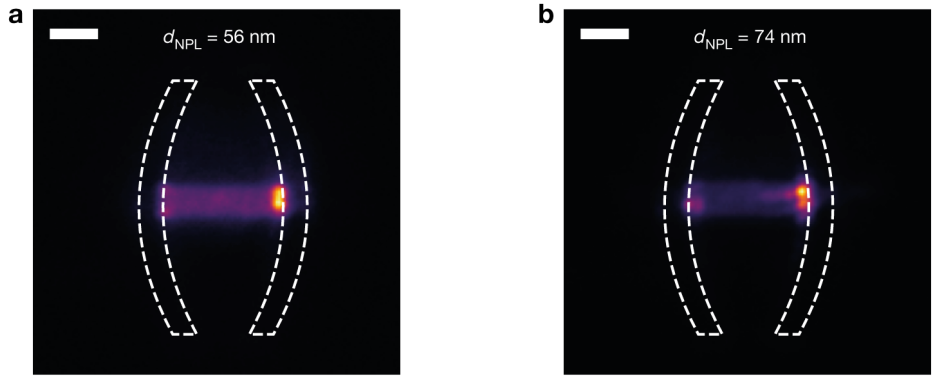

**Fig. S2. Metallic-cavity-laser images above threshold. a** Image of a device with a 56-nm-thick NPL layer pumped well above the lasing threshold (pump fluence 37 µJ/cm²). **b** Image of a device with a 74-nm-thick NPL layer pumped well above the lasing threshold (pump fluence 23 µJ/cm²). The devices in **a** and **b** lased in the plasmonic and photonic mode, respectively, and correspond to the devices probed to obtain the spectra in Fig. 2a–d and the light–light curves in Fig. 4a, b. The white dotted lines show the outline of the cavity reflectors. The colors are scaled to the maximum and minimum of the image. The scale bars are 4 µm.

**Section S3. Lasing condition**

At the lasing threshold, the electric-field amplitude, $E_0$, at an arbitrary point in the cavity returns to its original value after one cavity round trip[S3-5]:

$$E_0 = E_0\sqrt{R_1 R_2}\,e^{ik_z 2L_{cav}}. \tag{S1}$$

Here, $R_1 = R_2 = R$ is the mirror reflectivity, $k_z$ the propagation constant of the mode, and $L_{cav}$ the cavity length. This requirement can be split into a phase and amplitude condition:

$$m\frac{\lambda_0}{n_{\text{eff}}} = 2L_{\text{cav}}, \tag{S2}$$

$$G_{\text{mod}} \geq -\frac{\ln(R)}{L_{\text{cav}}}, \tag{S3}$$

where $m$ is an integer number, $n_{\text{eff}}$ the effective mode index, and $G_{\text{mod}} = -2k_z''$ the modal gain, which can be expressed in terms of the imaginary part of the propagation constant, $k_z''$. The phase condition (Eq. S2) requires integer multiples of the modal wavelength to fit into a round-trip cavity length, while the amplitude condition (Eq. S3) dictates the modal gain to be at least as large as the round-trip reflection losses. (Since we are not interested in a steady-state laser oscillation, the modal gain can also be larger than the round-trip reflection losses.) Note that we ignore scattering losses of the NPL stripe and only consider Ag losses (included in the modal gain) and reflection losses at the reflectors as possible loss channels (see SI Section S2).

**Section S4. Estimation of mirror reflectivity of Ag reflectors**

The mirror reflectivity can be estimated with the transfer-matrix method[S6,S7] applied to a three-layer stack with the first layer being an effective medium with the complex effective index of the plasmonic and photonic modes as refractive index, the second layer being a 10-nm-thick alumina film, and the third layer being Ag (Fig. S3a). For alumina and Ag, the refractive indices obtained by ellipsometry were used (see SI Section S13). The reflection and transmission coefficients used for the transfer-matrix method were calculated as described in ref. S8. The as-obtained mirror reflectivities are high (> 97%) for both the plasmonic and photonic modes (Fig. S3b). The actual reflectivities are lower than the calculated values, as scattering at the top of the 500-nm-tall reflectors is ignored[S9,S10]. Therefore, we assume a reflectivity of ~90% for both modes. For simplicity, we keep the reflectivity constant for all gain-layer thicknesses. As long as the reflectivity is high, the precise value is insignificant because the cavity losses are dominated by propagation losses (*i.e.* the ohmic losses in the Ag). A thorough discussion on the influence of the reflectivity on the reflection losses and the lasing mode is presented in SI Section S9.

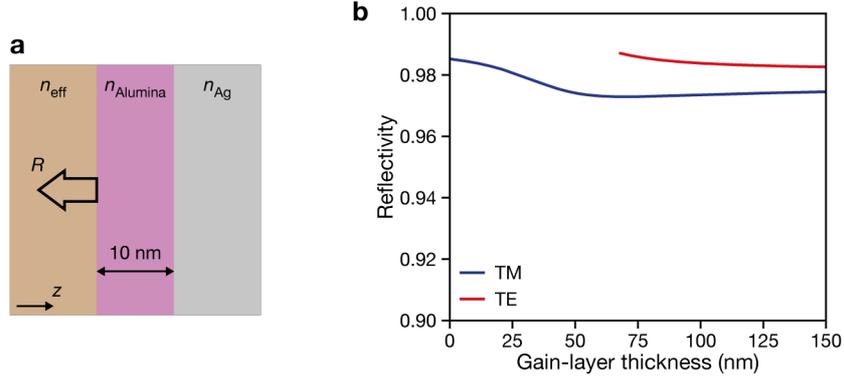

**Fig. S3. Estimation of the mirror reflectivity. a** Three-layer stack used to calculate the mirror reflectivity, $R$, using the transfer-matrix method. The first layer (brown) has the effective index of the TM and TE modes as refractive index, the second layer is a 10-nm-thick alumina film (purple), and the third layer is Ag (grey). The mode propagates in the $z$-direction towards the Ag layer. **b** Calculated reflectivity values as a function of gain-layer thickness for the three-layer stack in **a** for the plasmonic (TM, blue) and photonic (TE, red) modes. The true reflectivity values are lower, as scattering at the top of the 500-nm-tall reflectors is ignored in the transfer-matrix method.

## Section S5. Confinement factor, propagation length, and threshold gain

The confinement factor, $\Gamma$, is extracted from the linear relationship between the modal gain, $G_{mod}$, and the material gain, $G_{mat}$ (see Eq. 2 in the main text). The slope defines the confinement factor and the $y$-axis offset corresponds to the negative inverse of the propagation length, $L_{prop}$, for a transparent gain layer ($G_{mat} = 0$). We calculate the modal gain for various material gain values using the multilayer-waveguide model (Fig. S4a). Then, we fit Eq. 2 from the main text to the result to obtain the confinement factor and the propagation length. This procedure is repeated for a range of gain-layer thicknesses (Fig. S4b, c). The threshold gain is obtained through Eq. 4 in the main text (Fig. S4d). In principle, an analytical expression for the confinement factor in layered waveguide structures can be used to determine the confinement factor[S4,S11,S12]. However, extracting the confinement factor from a linear fit to the modal gain can also be employed for more complex structures (*i.e.* structures that are not invariant in the propagation direction) for which no analytical expression exists.

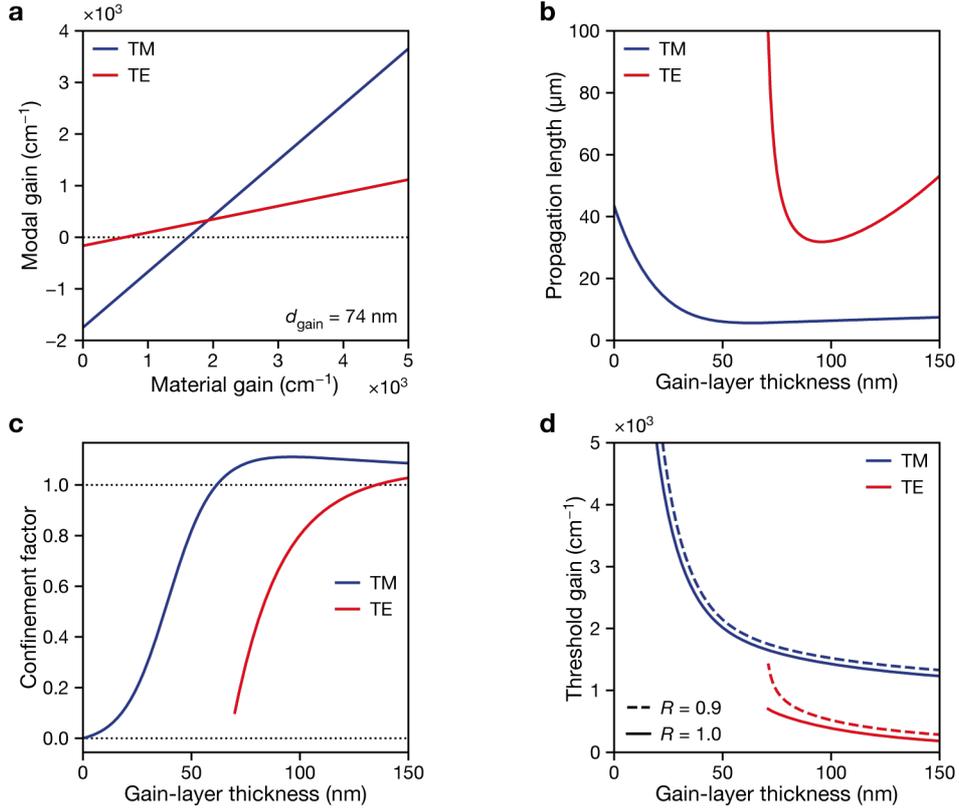

**Fig. S4. Calculated confinement factor and threshold gain. a** Modal gain as a function of material gain calculated for a waveguide structure with $d_{\text{gain}} = 74$ nm. The linear relationship is used to extract the confinement factor and the propagation length for the plasmonic (TM, blue) and photonic (TE, red) modes. **b** Propagation length as a function of gain-layer thickness. **c** Confinement factor as a function of gain-layer thickness. **d** Threshold gain as a function of gain-layer thickness. Here, the threshold gain was calculated without reflection losses ($R = 1.0$) and for a reflectivity $R = 0.9$.

**Section S6. Gap-layer design criteria**

As discussed in the main text, we find that for a HIG layer, the threshold gain for plasmonic lasing in devices with thin gain-layers is reduced compared to a LIG layer. Those conclusions were made for a multilayer-waveguide structure consisting of Ag at the bottom, a 10-nm-thick gap layer with a high or low refractive index, a NPL gain layer, and air on top. For the gain layer, the NPL refractive index at a vacuum wavelength $\lambda_0 = 635$ nm is $n_{\text{NPL}} = 1.89$ (see SI Section S13). The refractive index of bulk semiconductor materials, which are commonly used as gain media for nanolasers, is usually much larger. Therefore, we calculate the modal gain for a waveguide structure that employs a gain layer with a refractive index of $n_{\text{CdS}} = 2.5$ (representing bulk cadmium sulfide) at a free-space wavelength of $\lambda_0 = 635$ nm (Fig. S5a) and a structure with $n_{\text{InGaAs}} = 3.5$ (representing bulk indium gallium arsenide) at a free-space wavelength of $\lambda_0 = 1550$ nm (Fig. S5b). For the case of $n_{\text{CdS}} = 2.5$ and $\lambda_0 = 635$ nm, the threshold gain for gain-layer thicknesses slightly below the cutoff thickness of the photonic mode is reduced for the HIG structure, even if reflection losses are neglected ($R = 1.0$). For the case of $n_{InGaAs} = 3.5$ and $\lambda_0 = 1550$ nm, even for gain-layer thicknesses far above the photonic-mode cutoff thickness, the threshold gain is lower for the HIG structure.

Furthermore, we compare our hypothetical LIG and HIG structures (with the NPL refractive index for the gain layer and $\lambda_0 = 635$ nm) to the experimentally studied waveguide structure with an alumina gap layer ($n_{gap} = 1.64$) and to a structure without a gap layer (Fig. S5c). Avoiding a gap layer increases the overlap of the plasmonic mode (and also to a small extent the overlap of the photonic mode) with the gain layer, which reduces the threshold gain. For gain layers with $d_{gain} < 41$ nm, the threshold gain for the structure without gap layer (blue dotted line in Fig. S5c) starts to increase more steeply than for the other structures. This is a consequence of the overall waveguide structure being 10 nm thinner without the gap layer. The reduction in the overall thickness becomes also apparent for the photonic mode for which the cutoff thickness is shifted by 10 nm toward thicker gain layers. Even though removing the gap layer could potentially enable plasmonic lasing for very thin devices, it likely also compromises device lifetime, as exposure to ambient conditions can detrimentally affect the Ag quality.

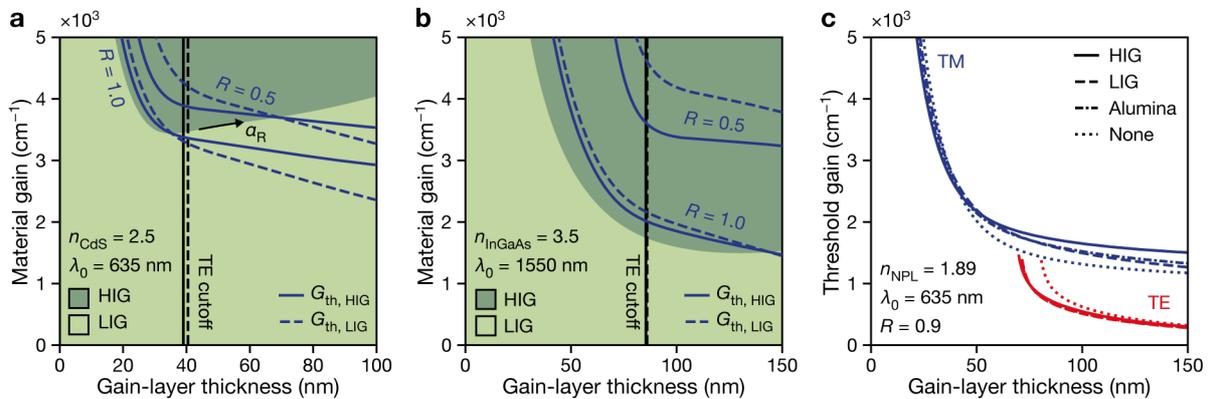

**Fig. S5. Threshold gain for various gain and gap layers. a, b** Material gain at which the modal gain is larger for the HIG (dark-green area) or LIG (light-green area) layer for a structure with a gain-layer refractive index of $n_{CdS} = 2.5$ and a free-space wavelength $\lambda_0 = 635$ nm (**a**), and $n_{InGaAs} = 3.5$ at $\lambda_0 = 1550$ nm (**b**). The threshold gains, $G_{th}$, for the plasmonic modes of the HIG and LIG structures are plotted as solid and dashed blue lines, respectively. They are calculated for a 10-µm-long cavity without reflection losses ($R = 1.0$) and for a mirror reflectivity $R = 0.5$. For increasing reflection losses, $\alpha_R$, the threshold-gain crossing point moves along the mode-condition border in the direction of the arrow (only shown in **a**). The black vertical solid and dashed lines indicate the photonic-mode (TE) cutoff thickness for the HIG and LIG structures, respectively. **c** Threshold gain for plasmonic modes (TM, blue) and photonic modes (TE, red) calculated for a NPL gain layer ($n_{NPL} = 1.89$) and various gap layers at $\lambda_0 = 635$ nm. The cavity length is 10 µm and the mirror reflectivity $R = 0.9$.

**Section S7. Inflection strength of light–light curve at threshold**

We can simplify the relationship between the pump power and the output-photon number to a regime above and below threshold. Therefore, we need to look through which channels the excited carrier population decays and to what fractions (see Eq. S5 and S6). Below threshold, the stimulated-emission term accounts for reabsorption (the material gain is negative) and thus does not contribute to a loss of carriers. Therefore, non-radiative decay and spontaneous emission are the only carrier-decay channels, summing to a total decay rate of $1/(\tau_{sp}\Phi)$, where $\tau_{sp}$ is the spontaneous-emission lifetime and $\Phi$ the quantum yield of the gain medium. The rate of carriers that feed into the surface plasmon or photon populations is described solely by the spontaneous-emission term $\beta_i/(\tau_{sp})$ with the spontaneous-emission factor $\beta_i$ of the mode $i$ ($i$ = TM, TE). Thus, the fraction of carriers that decay into a given mode population becomes $\beta_i\Phi$ below threshold. In contrast, above threshold, stimulated emission dominates and becomes the main decay channel for carriers. Therefore, we can assume that nearly all excited carriers decay into the lasing mode, resulting in a carrier-to-mode fraction of unity. Consequently, the difference in output-photon number below and above threshold (we can assume the output-photon number is proportional to the carrier-to-mode fraction) is on the order of $1 - \beta_i\Phi$, giving rise to a strong inflection in the light–light curve at the lasing threshold if $\beta_i\Phi$ is considerably smaller than unity.

**Section S8. Spontaneous-emission-factor calculation**

We estimate the fraction of spontaneous emission into a specific target mode, $\beta_i$, using the formalism described in ref. S13, providing exact solutions to Maxwell's equations for the power dissipated by an electric-dipole source in a multilayer structure. In brief, we treat the NPLs as randomly oriented electric-dipole sources in the gain layer of our multilayer-waveguide structure and calculate the position-averaged fraction of power dissipated into the target mode. The multilayer structure was modeled as a semi-infinite layer of Ag at the bottom, 10 nm of alumina, a gain layer of thickness $d_{gain}$, and a semi-infinite air layer on top, resulting in the same structure as used for waveguide-mode calculations described in the main text. The permittivities of Ag, alumina, and the gain layer (NPL layer) were extracted from ellipsometry data at a free-space wavelength $\lambda_0 = 635$ nm (see SI Section S13). Only the real part of the permittivity of the NPL film was retained in our approximation of $\beta_i$ to model spontaneous emission at excitation fluences just below the transparency of the gain medium. For a given gain-layer thickness, $d_{gain}$, we calculated the dissipated power as a function of the in-plane wavevector, $k_\parallel$, by an electric-dipole source at a height $x_0$ within the NPL layer (Fig. S6a). We considered the random orientation of the NPLs by averaging the power spectra obtained for dipole sources pointing along the two in-plane directions and the out-of-plane direction at a given height $x_0$. Spontaneous emission into the plasmonic and photonic modes are visible as peaks in the as-obtained $k_\parallel$-resolved dissipated-power spectra (Fig. S6b). Each peak is centered around the in-plane wavevector of the corresponding mode. As expected, the peak of the photonic mode is only visible for gain-layer thicknesses above the photonic-mode (TE) cutoff thickness (not shown). The fraction of spontaneous emission into mode $i$ ($i$ = TM, TE) for a randomly oriented dipole source at height $x_0$ within the gain layer was then extracted by integrating the dissipated-power spectrum over the peak corresponding to mode $i$ and dividing by the total dissipated power by emission into all modes, including emission of free-space photons and quenching in the Ag (integration of the dissipated power spectrum over all $k_\parallel$ from zero to infinity). Finally, $\beta_i$ was obtained for a given gain-layer thickness by varying $x_0$ from zero to $d_{gain}$ in steps of 2 nm and averaging over the obtained fractions.

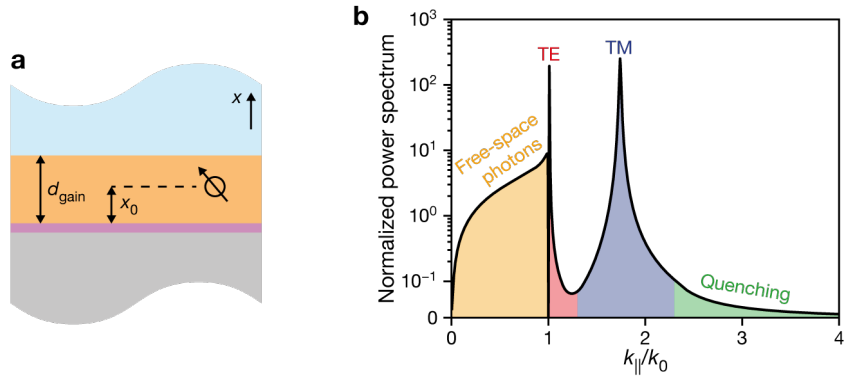

**Fig. S6. Calculation of the spontaneous-emission factor. a** Schematic of a dipole at height $x_0$ = 40 nm in a multilayer-waveguide structure with a NPL gain layer of thickness $d_{gain}$ = 74 nm. **b** Dissipated-power spectrum as a function of the in-plane wavevector, $k_\parallel$, normalized by the free-space wavevector, $k_0 = 2\pi/\lambda_0$, for the geometry in **a**. The spectrum is averaged over two in-plane directions and the out-of-plane direction of a dipole. The shaded areas indicate the range of integration for various modes: free-space photons (orange), photonic mode (TE, red), plasmonic mode (TM, blue), and quenched emission (green).

**Section S9. Effect of mirror reflectivity on lasing mode**

For the results in the main text, we have assumed equal mirror reflectivity for plasmonic and photonic modes. Here, we would like to explore how the reflection losses influence the threshold gain and how separate tuning of the modal reflectivities can potentially suppress the photonic mode in devices with gain-layer thicknesses above the photonic-mode (TE) cutoff thickness.

The reflection losses can be calculated using Eq. S3 (the losses are equal to the expression on the right-hand side). Both the cavity length and mirror reflectivity will influence the reflection losses when expressed as round-trip loss. For smaller cavities and lower mirror reflectivity the reflection losses increase (Fig. S7a). The influence of reflection losses on the threshold gain is inversely proportional to the confinement factor (see Eq. 4 in the main text). Hence, the threshold gain for the photonic mode strongly increases as its cutoff gain-layer thickness is approached (Fig. S7b). This now gives two possible tuning knobs for enabling plasmonic lasing in devices that support both plasmonic and photonic modes: selectively decreasing the reflectivity of the photonic mode or decreasing the reflectivity of both modes in a device operating slightly above the photonic-mode cutoff thickness.

Fig. S7c outlines for a given gain-layer thickness which mode experiences a lower threshold gain. Any combination of mode reflectivities below (above) a given gain-layer-thickness line will result in a lower threshold gain for the plasmonic (photonic) mode. For thick gain layers ($d_{gain}$ = 150 nm, black line in Fig. S7c), a low but equal reflectivity is not sufficient to lift the photonic threshold gain above the plasmonic threshold gain. In this case, the photonic-mode reflectivity must be substantially decreased compared to the plasmonic-mode reflectivity to achieve a lower threshold gain for the plasmonic mode.

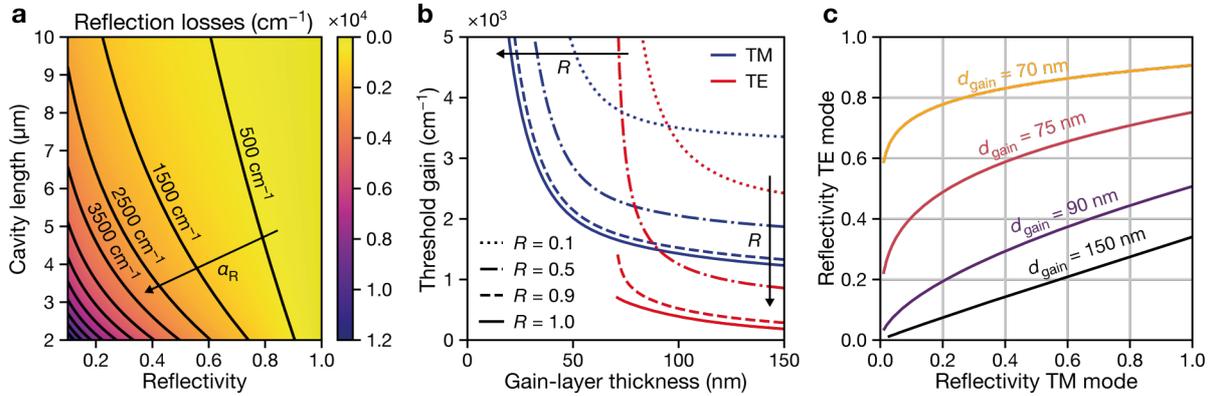

**Fig. S7. Influence of mirror reflectivity on lasing mode. a** Reflection losses, $\alpha_R$, arising from partially reflective mirrors are plotted as a function of mirror reflectivity and cavity length. The arrow points in the direction of increasing reflection losses. **b** Threshold gain for plasmonic (TM, blue lines) and photonic (TE, red lines) modes as a function of gain-layer thickness for a 10-μm-long cavity and various reflectivity values, $R$. Arrows indicate increasing mirror reflectivity. **c** Plasmonic (TM) and photonic (TE) mode reflectivity map. For a given gain-layer thickness (indicated by the line label), each point on the line indicates a combination of plasmonic and photonic mode reflectivities that results in the same threshold gain for both modes. A combination of plasmonic and photonic mode reflectivities below (above) a given gain-layer-thickness, results in a lower threshold gain for the plasmonic (photonic) mode. A cavity length of 10 μm was used for this calculation.

**Section S10. Fabrication of laser cavities**

The Ag cavities were fabricated by template stripping to achieve a smooth Ag surface[S14]. The template was prepared from a 1-mm-thick, 2-inch-diameter, <100> silicon wafer. The wafer was prepared with a 280-nm-thick positive electron-beam resist (Allresist, CSAR AR-P 6200.09). After electron-beam exposure (Vistec Lithography, EBPG 5200+), the resist was developed (Allresist, AR 600-546) for 1 min, revealing a negative mask of the cavity reflectors on top of the silicon wafer. The cavity length was 10 μm, the curved reflector is a parabola with a radius of curvature of 20 μm, the reflector width was 2 μm, and the reflector length was 21.2 μm. The curved shape of the reflectors leads to a stable Fabry–Pérot resonator[S3]. The wafer was etched ~500 nm deep using a hydrogen-bromide-based inductively-coupled-plasma reactive-ion etch (Oxford Instruments, Plasmalab System 100, 80 W radio-frequency power, 800 W inductively-coupled-plasma power, 50°C) for 150 s. The resist was removed by applying oxygen plasma (PVA TePla, GIGAbatch 310M) at 600 W for 5 min and subsequently dipping the silicon template into buffered hydrofluoric acid (Technic France, 1:7 hydrofluoric acid in ammonium fluoride) for 10 s. Finally, the template was cleaned in a Piranha solution [1:1 sulfuric acid (Sigma-Aldrich, 95.0–97.0%) and hydrogen peroxide (VWR Chemicals, 30%)] for 15 min.

Before Ag was deposited onto the template, it was cleaned in nitric acid (Sigma-Aldrich, ≥ 65%) for 15 min and rinsed with ultrapure water (18 MΩ cm) for ~3 min. Then, the template was sonicated in ultrapure water and isopropanol (Sigma-Aldrich, ≥ 99.5%) for 10 min each. A thermal evaporator (Kurt J. Lesker, Nano36) was used to deposit a ~700-nm-thick film of Ag (Kurt J. Lesker, 99.99%). The deposition was performed at a base pressure of $< 9 \times 10^{-8}$ mbar and at a deposition rate of 25 Å/s while the template was rotated at 60 rpm. A microscope slide was bonded to the Ag film using epoxy (Epoxy Technology, EPO-TEK OG116-31) cured under ultraviolet light for 2 h.

The Ag cavities were stripped manually from the template shortly before they were loaded into an atomic-layer deposition chamber (Picosun, Sunale R-150). A 10-nm-thick alumina layer was deposited at 50 °C applying 100 atomic-layer deposition cycles. Each cycle consisted of a 0.2 s trimethylaluminum pulse (Sigma Aldrich Fine Chemicals, electronic grade) at a flow rate of 200 sccm, a 10 s nitrogen purge, a 0.5 s water pulse (ultrapure, 18 MΩ cm) at a flow rate of 200 sccm, and a 30 s nitrogen purge.

For depositing the gain medium, CdSe/Cd$_x$Zn$_{1-x}$S core/shell NPLs having a 4-monolayer-thick CdSe core and a 2-nm-thick shell were synthesized according to ref. S15. The ink for the electrohydrodynamic nanoprinting was prepared by transferring the NPLs from hexane to tetradecane through selective evaporation while adjusting the concentration to an optical density of 5.0 (measured at the lowest-energy exciton peak using a quartz cuvette with a 10-mm path length). A description of the nanoprinting setup can be found elsewhere[S16]. 0.5 µL of the NPL ink was injected into the metal-coated nozzle with an outer diameter of ~1.7 µm. To eject ink, 250 V direct current were applied between the nozzle (+) and the indium-tin-oxide-coated-glass sample holder (ground). The distance between the nozzle and the metallic-cavity sample was kept at 5–10 µm. The stripes of 10 µm length and 2 µm width were generated by moving the sample stage in a serpentine-like fashion to print nine parallel lines at a pitch of 250 nm. Different stripe thicknesses were produced by varying the number of overprints.

To determine the required stage velocity and number of overprints for a given stripe thickness, a parameter sweep was performed using the same ink-loaded nozzle. Therefore, stripes were printed on a flat Ag–alumina substrate (which was fabricated in the same batch as the Ag-cavity sample) applying stage velocities from 4 to 16 µm/s, while the number of overprints was increased from one to ten. A dark-field reflection image (taken with a Nikon CFI LU Plan Fluor BD 50× objective with a numerical aperture of 0.8 on a Nikon Eclipse LV100 microscope) of the printed stripes was used to estimate the obtained stripe thickness for a given parameter combination by comparing the color of the stripe (the stripe waveguide can only accept certain input wavelengths) to a reference image of stripes whose thicknesses were determined by atomic force microscopy. Afterwards, stripes were printed into multiple cavities using the predefined stage velocities and overprint numbers to yield stripes with thicknesses between 20 and 90 nm. After all optical measurements were performed, the stripe thickness of each device was measured using atomic force microscopy (Bruker, Dimension FastScan). Scanning-electron micrographs of laser devices were acquired on a scanning-electron microscope (Hitachi, S-4800).

**Section S11. Optical characterization**

All optical measurements were performed in a closed-cycle helium cryostat (Montana Instruments, Cryostation 2 with LWD option) under vacuum and cooled to 4 K. The sample was mounted on a piezo-positioning system (Attocube, 1×ANPz101 and 2×ANPx101) inside the cryostat.

To obtain cold-cavity spectra, light from a 385-nm LED (Thorlabs, M385LP1) was used for wide-field excitation. The light from the LED was passed through a 400-nm/40-nm-bandwith bandpass filter (Thorlabs, FBH400-40), deflected at a 400-nm dichroic longpass filter (Omega Optical, 400DCLP), again deflected at a 405-nm dichroic longpass filter (AHF analysentechnik, F48-403), and imaged through the cryostat window onto the sample plane using a 60× extra-long-working-distance objective (Nikon, CFI S Plan Fluor ELWD with a numerical aperture of 0.7).

Lasing experiments were performed with a 405-nm pulsed laser source (~340 fs pulse duration, 1 kHz repetition rate) as excitation. The laser pulses emerged from a collinear optical parametric amplifier (Spectra-Physics, Spirit-OPA) pumped by a 1040-nm laser (Spectra-Physics, Spirit-1040-8). After spectral filtering, the beam was passed through a reflective, graduated neutral-density filter wheel to adjust the pulse power. Then, it was directed through a beam-expanding telescope and a half-wave plate (Thorlabs, AHWP05M-600). The beam passed through a 500-nm (Thorlabs, FESH0500) and a 750-nm shortpass filter (Thorlabs, FESH0750). Then, a fraction of the beam was deflected onto a photodiode (Thorlabs, S120VC) to monitor the pump power with a power meter (Thorlabs, PM100D). The rest of the beam was passed through a defocusing lens [achromatic doublet with focal length of 300 mm, (Thorlabs, AC254-300-A-ML)] and the 400-nm dichroic longpass filter (Omega Optical, 400DCLP), and then deflected at the 405-nm dichroic longpass filter (AHF analysentechnik, F48-403). Finally, the beam arrived at the 60× extra-long-working-distance objective from where it passed through the cryostat window and arrived at the sample plane as a defocused spot of ~30 μm diameter. Since the spectrometer slit appears in the image center, the excitation spot was aligned slightly off-center to conform with the center of the NPL stripe of the probed cavity.

The same 60× extra-long-working-distance objective was used for emission collection. The emission was directed through the 405-nm dichroic longpass filter (AHF analysentechnik, F48-403), a 450-nm longpass filter (Thorlabs, FEL0450) and relayed by six 200-mm focal-length achromatic doublets (Thorlabs, five AC254-200-A-ML and one AC508-200-A-ML) into an imaging spectrometer (Andor, Shamrock 303i) with the entrance slit set to 50 μm. Inside the spectrometer, the emission was horizontally dispersed by a 300 lines/mm grating (500-nm blaze) and imaged with an air-cooled electron-multiplying charged-coupled-device camera (Andor, iXon 888 Ultra). The sample was positioned such that the vertical entrance slit of the imaging spectrometer aligned onto the inner edge of one of the reflectors of the probed cavity. Each spectrum presented in the main text displays a single pixel row on the camera taken at the maximum intensity along the printed NPL stripe width (which roughly coincides with the vertical center of the stripe). One pixel row corresponds to a distance of ~0.22 μm along the reflector. For the integrated spectra in Fig. 5a and b, a single pixel row was integrated for each pump fluence. Integrating over multiple pixel rows did not significantly change the slopes (on the log–log scale) below and above the threshold. For real-space imaging the zero-order mode of the grating was used in combination with a wide-open slit. For polarization resolved spectra, a linear polarizer (Thorlabs, LPVISB100-MP2) was placed in an image-plane after the 450-nm longpass filter.

To determine the laser-spot diameter, the laser spot was imaged on a dried film of NPLs on the sample. Two perpendicular Gaussians were fitted through the beam center. The average of the distance between the 1/e-values of the Gaussians was taken as the beam diameter.

The pump power at the sample plane was measured using a photodiode (Thorlabs, S170C) in combination with a power meter (Thorlabs, PM100D). This allowed us to determine the conversion factor between the power reading of the photodiode that measured a fraction of the beam during the lasing experiments and the power arriving at the sample.

**Section S12. Multilayer-waveguide model**

The multilayer-waveguide model was implemented following a theoretical model described in ref. S17. In short, a general solution to the wave equation is taken, and the boundary conditions derived from Maxwell's equations are imposed at each layer boundary. In the top and bottom semi-infinite slabs, the wave is forced to exponentially decay, so only guided modes are accepted (no leaky modes)[S17,S18]. The resulting eigenvalue equation is solved by a minimization algorithm. We solve the eigenvalue equation for TM and TE modes. Within the considered range of gain-layer thicknesses (0–150 nm), only the fundamental TM and TE modes exist.

From the resulting propagation constant, $k_z$, we can derive the effective mode index, $n_{eff}$, the modal gain, $G_{mod}$, and the electric- and magnetic-field profiles. The time-averaged electromagnetic energy density, $\langle u_{EM} \rangle$, is calculated as defined by ref. S19:

$$\langle u_{EM} \rangle = \frac{1}{4}\varepsilon_0 \left( \varepsilon(\lambda) - \lambda \frac{\partial \varepsilon(\lambda)}{\partial \lambda} \right) |E|^2 + \frac{1}{4}\mu_0 |H|^2 \qquad (S4)$$

Here, $E$ and $H$ are the electric and magnetic fields, respectively, $\varepsilon$ and $\varepsilon_0$ are the relative and vacuum permittivities, respectively, $\mu_0$ is the vacuum permeability (all materials are assumed to be non-magnetic), and $\lambda$ denotes the wavelength. To get the total energy, $U_{EM}$, the electric- and magnetic-field intensities are integrated along the out-of-plane direction ($x$-direction) from minus to plus infinity. The time-averaged electric-energy density, $\langle u_E \rangle$, is calculated using only the electric-field term in Eq. S4.

As inputs to the model, the free-space wavelength, the permittivity of each constituent material, and the layer thicknesses needed to be defined. For the mode-profile and the modal-gain calculation the free-space wavelength, $\lambda_0$, was kept at $\lambda_0 = 635$ nm (1.95 eV). For the FSR calculations, $\lambda_0$ was varied over a range of 500–700 nm (1.48–1.77 eV). The relative permittivities were obtained from ellipsometry (see SI Section S13). The alumina layer was considered as lossless [Im($\varepsilon_{gap}$) = 0], while the Ag losses were included as a positive imaginary part of the relative permittivity [Im($\varepsilon_{Ag}$) > 0]. The gain layer was kept lossless for the calculation of the mode profiles and the mode index [Im($\varepsilon_{gain}$) = 0], lossy (as obtained from ellipsometry of a NPL film) for the FSR calculation, and with a positive material gain for the modal-gain calculations [Im($\varepsilon_{gain}$) < 0].

**Section S13. Ellipsometry of NPL, Ag, and alumina films**

To obtain the relative permittivity of the gain layer for the multilayer-waveguide calculations, ellipsometry was performed on a film of NPLs. NPLs dispersed in hexane with an optical density of 2.0 (measured at the lowest energy exciton peak using a quartz cuvette with a 10-mm path length) were mixed with octane (Sigma-Aldrich, 98%) in an 8:1 volumetric ratio. Then, 35 μL of the mixture was dropcast on a 20 mm × 20 mm, <100> silicon chip. A reference silicon chip of the same wafer was kept for determining the thickness of the native-oxide layer. Ellipsometry was performed on a J.A. Woollam VASE ellipsometer. Measurements were collected at four angles (60°, 65°, 70°, and 75°) in the wavelength range 300–1000 nm in steps of 1 nm in reflection mode. A Kramers–Kronig-consistent dispersion model was fitted to the obtained data. The film thickness was fitted together with a Tauc–Lorentz oscillator in the transparent spectral region of 700–1000 nm. Then, the spectral region was successively increased towards the blue wavelength regime, while in total five Gaussian oscillators were added to conform with the excitonic features at the band edge. The mean squared error of the final fit was 2.2. The fitted NPL-film thickness was confirmed with a cross-sectional scanning-electron-microscope image. From the dispersion model, the complex refractive index was extracted (Fig. S8a), and the relative permittivity was then derived from these data. To mimic the blue shift of the NPL bandgap when cooled to cryogenic temperatures, the refractive index values were shifted by 18 nm to the blue, corresponding to the shift in the photoluminescence spectrum. The complex relative permittivity, $\varepsilon_{gain}$, was then obtained through $\varepsilon_{gain} = (n + ik)^2$, with $n$ and $k$ being the real and imaginary part of the complex refractive index and i being the imaginary unit. (Note that for the modal gain calculations, the real part of the relative permittivity was set to $\varepsilon'_{gain} = n^2$, while the imaginary part was swept over a range of values obtained from Eq. 7 in the main text.)

To determine the relative permittivity of Ag, a Ag film was template-stripped from a flat region of a silicon template immediately before an ellipsometry scan in reflection mode was performed (angles 65°, 70°, 75°; scan range 400–1700 nm in steps of 10 nm). The optical constants were obtained by a numerical fit to the experimental data (Fig. S8b), and the complex relative permittivity, $\varepsilon_{Ag}$, was derived by employing $\varepsilon_{Ag} = (n + ik)^2$.

For acquiring the permittivity values of alumina, an alumina film was prepared by atomic-layer deposition onto a <100> silicon chip, using the same deposition parameters as for the device fabrication (see SI Section S10). A reference silicon chip of the same wafer was used to determine the thickness of the native-oxide layer. An ellipsometry scan in reflection mode was performed shortly after the deposition (angles 65°, 70°, 75°; scan range 400–900 nm in steps of 10 nm). Then, a transparent Cauchy model was fitted to the experimental data, the refractive index was extracted (Fig. S8c), and the relative permittivity was calculated using $\varepsilon_{gap} = n^2$.

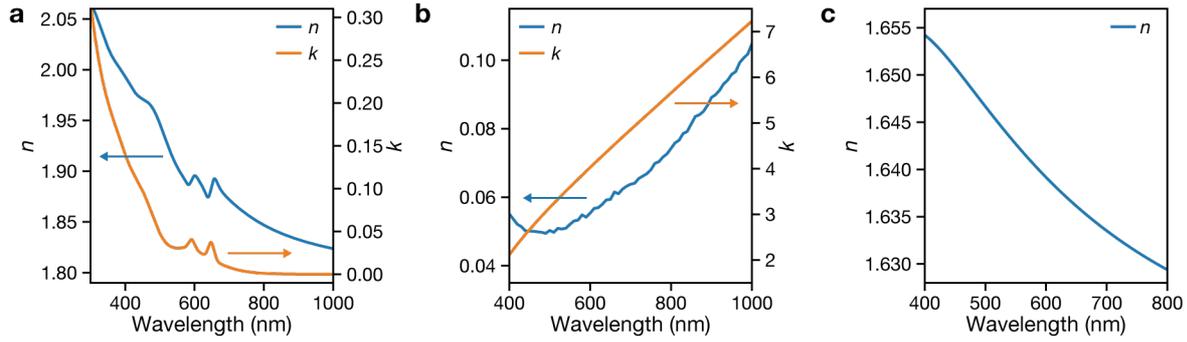

**Fig. S8. Complex refractive index of NPL, Ag, and alumina films. a** Complex refractive index ($\tilde{n} = n + ik$) of a dropcast NPL film obtained with ellipsometry in reflection mode. A Kramers-Kronig-consistent dispersion model comprising a Tauc–Lorentz and five Gaussian oscillators were fitted to the ellipsometry data. The mean squared error of the fit was 2.2. The data displayed here do not include the 18-nm blue shift to imitate the bandgap shift at cryogenic temperatures. **b** The Ag refractive index was obtained with ellipsometry performed on a flat, template-stripped Ag film. Through numerical fitting of the experimental data, the complex refractive index was obtained. **c** An alumina film was prepared in the same manner as for the device fabrication but on a silicon substrate. Ellipsometry data was acquired and a transparent Cauchy model was fitted to extract the refractive index.

**Section S14. Waveguide dispersion**

The dispersion of the effective index was calculated by sweeping over the free-space wavelength in the multilayer-waveguide model. The waveguide formed by the gain layer strongly influences the effective mode index (Fig. S9a). The mode index is highly dispersive (Fig. S9b), not only because the constituent materials are dispersive, but also because the gain layer forms a waveguide that introduces waveguide dispersion. As a consequence, the group index takes large values (Fig. S9c).

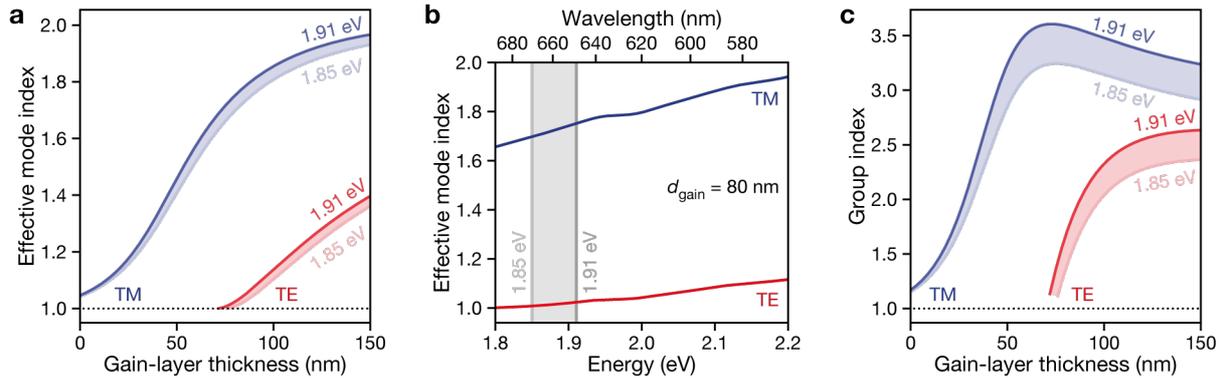

**Fig. S9. Effective index dispersion. a** Effective mode index in the range 1.85–1.91 eV (lines and shaded areas) for the plasmonic (TM, blue) and photonic (TE, red) mode. **b** Dispersion of the effective mode index for a gain-layer thickness of 80 nm for the plasmonic (TM, blue) and photonic (TE, red) mode. The grey shaded region indicates the range 1.85–1.91 eV. **c** Group index in the range 1.85–1.91 eV (lines and shaded areas) for the plasmonic (TM, blue) and photonic (TE, red) mode as a function of gain-layer thickness.

**Section S15. Laser rate equations**

The coupled rate-equation model was set up following refs. S6, S20:

$$\frac{dN}{dt} = \frac{\eta_P P(t)}{E_P} - \frac{N}{\tau_{sp}} - k_1 N - \sum_i v_{g,i} \Gamma_i G_{mat}(N) S_i, \quad (S5)$$

$$\frac{dS_i}{dt} = v_{g,i}\left(\Gamma_i G_{mat}(N) - \frac{1}{L_{prop,i}} + \frac{\ln(R)}{L_{cav}}\right) S_i + \beta_i \frac{N}{\tau_{sp}}, \quad (S6)$$

with $t$ being the time, $N$ the excited-carrier population, $S_i$ the surface-plasmon and photon populations, respectively, with $i$ indicating the mode ($i$ = TM, TE). For laser cavities with gain-layer thickness above the photonic-mode-cutoff thickness, a surface-plasmon, $S_{TM}$, and a photon, $S_{TE}$, mode population were considered, while for laser cavities with gain-layer thicknesses below the photonic-mode cutoff thickness, only a surface-plasmon mode population, $S_{TM}$, was taken into account. $\eta_P$ is the fraction of pump power absorbed by the gain layer (it is estimated using the transfer matrix method; see SI Section S16), $E_P$ the energy of the pump photon (at 405 nm), and $P(t)$ the optical pump power. The time dependence of the optical pump power is modeled as a squared hyperbolic secant function[S21]: $P(t) = P_P \text{sech}^2(1.76t/\Delta t)$, with the peak power $P_P = 0.88\, E_{pulse}\Delta t$, the energy of the pump pulse $E_{pulse}$, and the pulse duration $\Delta t$ = 340 fs. $\tau_{sp}$ and $k_1$ are the spontaneous-emission lifetime and the non-radiative recombination rate, respectively. Both are estimated from the total lifetime $\tau_{tot}$ = 5 ns and the quantum yield of our NPLs in solution, $\Phi$ = 0.88 (ref. S15), as $\tau_{sp} = \tau_{tot}/\Phi$ and $k_1 = 1/\tau_{tot} - 1/\tau_{sp}$. Note, that we do not include any Auger-recombination term in the rate equations, as evidence suggests that Auger processes in NPLs are suppressed at cryogenic temperatures[S22]. The group velocity of mode $i$ is defined through $v_{g,i} = c/n_{g,i}$, where $n_{g,i}$ is the group index of mode $i$ (defined in Eq. 6 in the main text). The confinement factor $\Gamma_i$ and the propagation length $L_{prop,i}$ were obtained from the modal-gain calculations (see SI Section S5). The material gain $G_{mat}(N)$ for the NPL layer was modeled with a logarithmic function with parameters fitted from data in ref. S23 (see SI Section S17). The mirror reflectivity $R = 0.9$ and the cavity length $L_{cav}$ = 10 μm were kept the same for the plasmonic and photonic modes. $\beta_i$ denotes the spontaneous-emission factor and was estimated by calculating the branching ratio of an orientation- and position-averaged electric dipole in the multilayer-waveguide structure (see SI Section S8).

The mirror loss rate, defined as $v_{g,i}\ln(R)/L_{cav}S_i$, was integrated over time to obtain the output-photon number. Here, it is assumed that all surface-plasmons and photons lost through imperfect mirror reflection contribute to the output-photon number. In our experiments, only a portion of the surface-plasmons and photons scatter to photons within the collection cone of our objective. We assume a linear relation between the portion of collected light and electromagnetic intensity inside the cavity modes. This allows us to compare the output-photon number to integrated-intensity values from experiments.

Further, we ignore modal cross-coupling and the Purcell enhancement in our rate-eqution model. As the field components of the plasmonic and photonic mode are orthogonal to each other, no direct cross-coupling is expected. Thus, no cross-coupling terms between the surface-plasmon and photon populations are included. The surface-plasmon and photon populations indirectly interact with each other through the carrier population. The Purcell enhancement is neglected in the rate-equation model, as it has been shown that, in cavities of this size, the Purcell effect plays a minor role[S24].

**Section S16. Fraction of absorbed pump power**

The power of the pump beam absorbed by the gain layer was estimated using the transfer-matrix method described in ref. S7. The matrix method was applied on the same waveguide structure as used for the multilayer-waveguide model. The reflection and transmission coefficients used for the transfer-matrix method were calculated as described in ref. S8. For the material parameters, the relative permittivities obtained by ellipsometry were used at the pump free-space wavelength $\lambda_0 = 405$ nm (see SI Section S13). For the gain layer, the complex permittivity of the NPL film was employed. The absorbed power was defined as the remaining power after subtracting the reflected power (at the air–gain-layer interface) and the transmitted power (into the Ag layer). Furthermore, the fraction of power incident on the NPL stripe was calculated by integrating a two-dimensional Gaussian over the width (2 µm) and length (10 µm) of the stripe centered to the Gaussian.

**Section S17. Material gain estimation for rate-equation model**

For the rate-equation model, the material gain as a function of excited carrier number was estimated for a NPL gain layer. A logarithmic model, commonly used to describe gain in semiconductor materials[S6], was fit to experimental data from ref. S23:

$$G_{\text{int}}(\langle N \rangle) = G_{\text{int},0} \ln\left(\frac{\langle N \rangle + N_s}{N_{tr} + N_s}\right). \tag{S7}$$

S19

Here, the intrinsic gain, $G_i$, is described as a function of the average carrier number per NPL, $\langle N \rangle$. The fit resulted in an empirical intrinsic gain coefficient $G_{int,0} = 22329$ cm$^{-1}$, an average transparency carrier number $N_{tr} = 4.9$, and an average shift carrier number $N_s = 51.8$ [which forces the natural logarithm to be finite for $\langle N \rangle = 0$]. To obtain the material gain, the intrinsic gain was multiplied by the NPL packing fraction, which was taken to be $f_{pack} = 40\%$ (ref. S25). Furthermore, the average carrier number was expressed as the total carrier number, $N$, divided by the total number of NPLs, $N_{NPL}$. The total number of NPLs was determined through $N_{NPL} = f_{pack} V_{stripe}/V_{NPL}$, with a stripe volume of $V_{stripe} = 10$ μm × 2 μm × $d_{NPL}$ (where $d_{NPL}$ is the NPL stripe thickness) and a NPL volume of $V_{NPL} = 22$ nm × 22 nm × 10 nm (estimated from scanning-electron-microscope images). The resulting function for the material gain could then be expressed as:

$$G_{\text{mat}}(N) = G_{int,0} \ln\left(\frac{\frac{N}{N_{NPL}} + N_s}{N_{tr} + N_s}\right) f_{pack}. \tag{S8}$$

**Section S18. Mode-switching discussion**

Our rate-equation model predicts that metallic-cavity lasers could be employed for switching between photonic and plasmonic lasing at high speeds within the same device. Both the photonic and plasmonic laser pulses evolve on a picosecond to sub-picosecond time scale (Fig. 5d). Note that this simplified rate-equation model does not account for the time for carriers to thermalize to the band edge of the NPLs. However, experiments have shown that this thermalization can occur on (sub-)picosecond timescales[S26,S27]. For simultaneous plasmonic and photonic lasing to occur, a gain medium that can provide a material gain of approximately 3000 cm$^{-1}$ is required. Even though our NPLs provide sufficient gain for plasmonic lasing for cavities with NPL-stripe thicknesses below the photonic-mode cutoff thickness, the devices with $d_{NPL} \geq 74$ nm displayed signs of degradation when the pump fluence was increased far beyond the photonic lasing threshold. We hypothesize that thick gain layers provide a worse heat sink than thin gain layers, and therefore limit the achievable material gain to lower values.

## Section S19. Supplementary references